\documentclass[prd,amsmath,amssymb,aps,superscriptaddress,10pt]{revtex4-2}
\usepackage[T1]{fontenc}
\usepackage[utf8]{inputenc}
\usepackage{mathtools}
\usepackage{graphicx}
\usepackage{caption}
\usepackage{subcaption}
\usepackage{xcolor}
\usepackage{float}
\usepackage{soul}
\usepackage{lipsum}

\usepackage{array} 
\usepackage[colorlinks=true, pdfstartview=FitV, linkcolor=blue, citecolor=blue, urlcolor=magenta, breaklinks=true]{hyperref}
\usepackage{orcidlink}
\usepackage{booktabs}
\usepackage{comment}

\begin{document}

\title{Bayesian approach study of hybrid neutron stars}
\author{F\'abio K\"opp \orcidlink{0000-0001-9970-4339}  }
\email{fabiokopp@proton.me}
\affiliation{Departamento de F\'{\i}sica - CFM, Universidade Federal de Santa Catarina, 88.040-900, Florian\'opolis/SC, Brazil}
\author{César H. Lenzi \orcidlink{0000-0001-5887-338X}}
\email{chlenzi@ita.br}
\affiliation{Departamento de Física e Laboratório de Computação Científica Avançada e Modelamento (Lab-CCAM),
Instituto Tecnológico de Aeronáutica, DCTA, 12228-900, São José dos Campos/SP, Brazil}

\author{César V. Flores \orcidlink{0000-0003-1298-9920}}
\email{cesar.vasquez@uemasul.edu.br}
\affiliation{Centro de Ciências Exatas, Naturais e Tecnológicas, Universidade Estadual da Região Tocantina do Maranhão, 65901-480, Imperatriz/MA, Brazil}
\affiliation{Programa de Pós-graduação em Física - CCET,
Universidade Federal do Maranhão,
65080-805, São Luís, Maranhão, Brazil.}

\author{D\'ebora P. Menezes \orcidlink{0000-0003-0730-6689}}
\email{debora.p.m@ufsc.br}
\affiliation{Departamento de F\'{\i}sica - CFM, Universidade Federal de Santa Catarina, 88.040-900, Florian\'opolis/SC, Brazil}

\begin{abstract}

In this work, we explore how astronomical observations (specifically measurements of masses and radii) can constrain the presence of quark matter inside neutron stars, namely the phase transition from nuclear matter to deconfined quark matter. Our approach employs Bayesian analysis to study this constraint. In this sense to model Hadronic matter we used the relativistic mean-field (RMF) approximation, for which we have selected two parameter sets: \(NL3^{*}\omega\rho\), representing hadronic matter with nucleons only, and $EL3\omega\rho$ with nucleons only and $EL3\omega\rho Y$, which includes hyperons. On the other hand deconfined quark matter is modeled using the vector-MIT bag model. Also, for our purpose, the first order phase transition is implemented using the Maxwell construction. Bayesian inference is performed by optimizing three parameters:
the bag constant (i.e. $B^{1/4}$), the vector coupling constant \(\left(G_{v}\right)\), and the Dirac sea contribution ($b_{4}$). 
We found that a phase transition could exist at densities below \(2.0\,n_{0}\) for both the $EL3\omega\rho $ and $NL3^{*}\omega\rho$ parametrizations. As a consequence, The inferred parameters of the vector MIT bag model, based on our likelihood function, imply hybrid stars with large quark cores.

\end{abstract}

\maketitle

\section{Introduction} 

A neutron star is one of the most fascinating astrophysical objects in the Universe. This compact object is well known for hosting some of the most extreme conditions in nature; for example it exhibits intense magnetic fields, high rotation speeds, intense gravitational forces and extreme baryonic densities at its center.
At these extreme densities, it is theorized that hadronic nuclear matter could undergo a phase transition to a deconfined phase of strange quark matter, making the existence of hybrid stars possible. In this respect, it is of vital importance to study the astrophysical signatures that could reveal the presence of quark matter at the interior of these compact objects.

Advanced terrestrial and space-based instruments are already making significant contributions to the study of compact stars. In particular, the LIGO/Virgo Collaboration has detected gravitational waves from neutron star mergers \cite{Abbott2018_GW170817_Radii}, and the Neutron Star Interior Composition Explorer (NICER) has observed neutron star properties through soft x-ray timing \cite{riley2021nicer,Miller2019,Miller2021}. These observatories now provide high-precision data on fundamental neutron star characteristics, such as mass, radius, moment of inertia, and tidal deformability. In fact such data are of outstanding importance for distinguishing between different types of compact stars: hadronic stars, pure quark stars, and hybrid stars.

It is well established that quark matter must be described by Quantum Chromodynamics (QCD), the fundamental theory governing strong interactions. However, an ab initio description of dense matter, whether hadronic or quarkyonic, cannot yet be derived directly from QCD. As a result, the quest for an effective model that can accurately represent the neutron star equation of state (EOS) remains a complex and active area of research, with numerous approaches proposed in the literature. To constrain the modeling of dense matter EOS, researchers employ various strategies, including experimental data on symmetric nuclear matter, results from lattice QCD and chiral effective field theory, as well as model-independent EOS fitting techniques \cite{Mariani:2024gqi}.

A similar study  was conducted in \cite{Kopp:2025hgp}, using the same parametrization for the hadronic matter ($NL3^{*}\omega\rho$) and the same model for the quark core (vMIT). However, that work did not incorporate the updated values from NICER observations, the perturbative Quantum Chromodynamics (pQCD) constraints, and the analysis was carried out heuristically.

In this context, Bayesian methods have become increasingly popular for exploring the parameter space of high-density nuclear matter models \cite{Imam:2021dbe,Malik:2022ilb,Coughlin:2019kqf}, as they provide a powerful framework for analyzing neutron star equations of state. These methods combine prior knowledge with observational data to generate probabilistic constraints on the parameters of the EOS. For example,  for the vector MIT bag model these parameters are ~( $B^{1/4}, G_{v}, b_{4}$). 

In this work, the deconfined quark phase is modeled using the vector MIT bag model, which is discussed in more detail in Sec. II. In the same section, the phase transition from hadronic to quark matter is treated using the Maxwell construction ~\cite{1992PhRvD}.
This approach not only refines viable EOS models, but also quantifies the uncertainties in their parameters. In Sec. III we present the stellar structure and the tidal deformability equations.

In Sec. II, we describe the microphysics of hybrid stars. In Sec. III, we discuss the macrophysics, i.e., how we calculate the mass, radius, approximate universal relations, dimensionless tidal deformability, and the $\Lambda_1$--$\Lambda_2$ relation. The Bayesian methodology is presented in Sec. IV. The results are presented in Sec. V. Finally, our conclusions are given in Sec. VI.

\section{Microphysics}

As mentioned in the introduction, a hybrid star consists of a core of  quark matter surrounded by an outer layer of hadronic matter. Due to the uncertain nature of matter at extremely high densities, the EOS is typically derived using phenomenological models that aim to capture the expected properties of matter across different density regimes. To this end, the deconfined quark matter in the core can be described using the vector MIT bag model (vMIT) \cite{Lopes:2020btp}, while the hadronic outer layer is modeled with the relativistic mean field (RMF) hadronic model \cite{Boguta1977}. The outer crust is represented by the BPS EOS \cite{bps}. This section provides a detailed overview of the theoretical microphysical framework used to construct these models.

\subsection{Hadronic Matter}

Hadronic matter is described using the relativistic mean field hadronic model 
\cite{walecka}, in which the strong interaction between hadrons is mediated by the exchange of virtual mesons. At the Lagrangian density level, the model is expressed as
\begin{align}
     \mathcal{L}_{\rm RMF}={}& \sum_{b}\overline{\psi}_b \left[\gamma_\mu\left(i\partial^\mu - g_{b \omega}\omega^\mu - g_{b \rho} \frac{\vec{\tau}_b}{2} \vec{\rho}^\mu \right)
    -m^*_b\right]\psi_b \nonumber \\
    &+\frac{1}{2}\partial_\mu \sigma \partial^\mu \sigma - \frac{1}{2} m_\sigma^2 \sigma^2 - \frac{1}{3!} \kappa \sigma^3 - \frac{1}{4!} \lambda \sigma^4 \nonumber \\
    & -\frac{1}{4} \Omega^{\mu\nu}\Omega_{\mu\nu} + \frac{1}{2}m_\omega^2 \omega_\mu \omega^\mu 
     - \frac{1}{4} \vec{R}_{\mu \nu} \vec{R}^{\mu \nu} 
     \nonumber \\
    &+ \frac{1}{2}m_\rho^2 \vec{\rho}_\mu \vec{\rho}^\mu + \Lambda_v g_{N \omega}^2 g_{N \rho}^2\omega_\mu \omega^\mu \vec{\rho}_\mu \vec{\rho}^\mu, 
     \label{nlwm}  
\end{align}
{where the Dirac spinor $\psi_b$ represents the baryon $b$ with the effective mass $m_b^*=m_b-g_{b \sigma} \sigma$, $\vec{\tau}_b$ are the corresponding Pauli matrices. The coupling constants of the mesons $i=\sigma,\, \omega,\text{ and} ~ \rho$ with the baryon $b$ are represented by $g_{bi}$, $m_i$ is the mass of the meson $i$, $\kappa$ and $\lambda$ are scalar self-interaction constants and $\Lambda_v$ is the $\omega-\rho$ crossing interaction coefficient. The $b$ sum extends over the baryons {\textbf{\(n, p , \Lambda, \Sigma^-, \Sigma^0, \Sigma^+, \Xi^-, \text{and}~ \Xi^0 \)}}.

The meson field tensors are $\Omega_{\mu \nu}=\partial_\mu \omega_\nu - \partial_\nu \omega_\mu$, $\vec{R}_{\mu \nu}=\partial_\mu\vec{\rho}_\nu - \partial_\nu\vec{\rho}_\mu - g_\rho(\vec{\rho}_\mu \times \vec{\rho}_\nu)$ and $\Phi_{\mu \nu}=\partial_\mu \phi_\nu - \partial_\nu \phi_\mu$. 
The parameter sets employed in this work are given in Table \ref{tab:parametros}, and the nuclear matter saturation properties are given in Table \ref{tab:propriedades}.

This hadronic model is fitted to finite symmetric nuclear matter at saturation density $(n_{0})$,
satisfying the following bulk properties: the binding energy $E/A$, compressibility modulus $K_0$, symmetry energy $S_0$, and its slope $L_0$. 
The parametrizations employed in this study are the NL3$^{*}\omega\rho$ \cite{Lopes:2021yga} and the EL3$\omega\rho$  \cite{Lopes:2021zfe}, 
both sets satisfying the constraints imposed by microscopic calculations of 
 nuclear matter \cite{dutra2014}. The presence of hyperons is also explored using the EL$3\omega\rho Y$ parametrization. The couplings of hyperons to vector mesons as well as scalar mesons are presented in the Appendix. This parametrization includes the strange isoscalar-vector field $\phi^{\mu}$, which enables the existence of hyperonic stars with masses of approximately $2 ~ \mbox{M}_{\odot}$. It is important to emphasize that  the EOSs used in this work are both in beta equilibrium and under local charge neutrality \cite{Lopes:2021yga, Lopes:2021zfe}.

\begin{table*}[!htb]
\caption{Parameter sets for the models discussed in the text. The meson masses $m_\sigma$, $m_\omega$, and $m_\rho$ are all given in MeV. The nucleon mass is $M=939$ MeV.}\label{tab:parametros}
\begin{tabular}{l|ccccccccc}
\hline\hline
 Model  & $m_\sigma$  & $m_\omega$  & $m_\rho$  & $g_{N \sigma}$  & $g_{N \omega}$  & $g_{N \rho}$  & $\kappa/g_\sigma^3$  & $\lambda/g_\sigma^4$  & $\Lambda_v$  \\ \hline
   NL3$^*\omega \rho$ & $502.574$  & $782.600$  & $763.000$  & $10.0944$  & $12.8065$  & $14.4410$  & $0.004417$  & $-0.017422$  & $0.045$ \\
  EL3$\omega \rho$ & $512.000$  & $783.000$  & $770.000$  & $9.0286$  & $10.5970$  & $9.4381$  & $0.008280 $  & $-0.023400$  & $0.0283$  \\ \hline\hline
\end{tabular}
\end{table*}

\subsection{Strange Quark Matter}

An important consideration in the high density regime is the strange matter hypothesis, which 
suggests that strange quark matter, 
 composed of up, down, and strange quarks may be
the true ground state of hadronic matter \cite{Bodmer:1971we,Terazawa:1989iw,Witten:1984rs}.

Studying matter under  extreme conditions is challenging due to the complexity of QCD. The two main approaches - lattice QCD (LQCD) and effective models - each have their limitations. For example, LQCD encounters difficulties such as the sign problem and limited applicability at low chemical potentials \cite{Nagata:2021ugx}, which makes it unsuitable for investigating high-density matter. Consequently, effective models are commonly used to study matter at high densities, particularly in compact objects like neutron stars. The limitations of relativistic mean-field models, for example, are associated with the requirement that $r_{0} \ll m_{\text{fields}}^{-1}$, where 
$r_{0}$ is the unit radius $\sim$ 1.2 fm and 
$m_{\text{fields}}$ denotes the meson masses. For instance, the $\rho$ meson has a mass of 775.26 MeV, corresponding to a length scale $hc/m  \approx0.254$  fm.
This condition is, however, completely violated both in nuclei and in neutron stars \cite{Heiselberg:2000dn}. 

The MIT bag model describes baryons as systems composed of three confined, non-interacting quarks enclosed within a finite spatial region known as the “bag.” In this framework, the bag is represented as an infinite potential well that ensures quark confinement~\cite{Chodos1974}.Quarks are treated as quasi-free particles inside the bag, moving relativistically within the confined volume while being forbidden from escaping through the boundary. The stability of the system arises from the balance between the outward pressure generated by the kinetic energy of the confined quarks and the inward vacuum pressure associated with the bag constant, which effectively maintains confinement. Thus, the model captures two key properties of QCD: asymptotic freedom and confinement.
In this case the MIT Lagrangian density is given by:
\begin{equation}
\mathcal{L_{\rm MIT}} = \sum_{u,d,s}\{ \bar{\psi}_q  [ i\gamma^{\mu} \partial_\mu - m_q ]\psi_q - B \}\Theta(\bar{\psi}_q\psi_q), \label{e1}
\end{equation}   
where $m_q$ is the $q$ quark mass, $\psi_q$ is the Dirac quark field, $B$ is the constant vacuum pressure and $\Theta(\bar{\psi}_q\psi_q)$ is the Heaviside step function that is included to assure that the quarks exist only 
inside the bag. 

In this work, we use a variant of the MIT bag model to describe the EOS of deconfined strange quark matter, know as the vector MIT bag model. In this formulation, quarks interact via a repulsive vector field analogous to the $\omega$  meson exchange in relativistic mean-field hadronic models. The lagrangian for this vector interaction is given by

\begin{equation}
\mathcal{L}_V = \sum_{u,d,s}g_{V}\{ \bar{\psi}_q  [ \gamma^{\mu} V_\mu  ]\psi_q \}\Theta(\bar{\psi}_q\psi_q), \label{v0}
\end{equation}
we also consider the mass term and the Dirac sea contribution\cite{Lopes:2020btp},
\begin{equation}
\mathcal{L}_V = \frac{1}{2}m_V^2V_\mu V^\mu + b_4 \frac{(g_{V}^{2} V_\mu V^\mu)^2}{4}, \label{v2}
\end{equation}
where the mass of the vector field, $m_V$, is taken to be 780 MeV, and $b_4$ is a dimensionless parameter to modulate the Dirac sea contribution. 

In this work, the strength of the vector channel is directly related
to $(g_{V}/m_V)^2$, so we define 
\begin{equation}
 G_V  = \bigg ( \frac{g_{V}}{m_V} \bigg )^2, \label{definition}
\end{equation}
and adopt $m_u=m_d=4$ MeV and $m_s=93$ MeV.

To ensure that strange matter remains electrically neutral and in chemical equilibrium, the following conditions must be satisfied\cite{Lopes:2020btp}. The effective chemical potential of the quarks $\mu_{q}$ and leptons ($l$) are: 
 \begin{eqnarray}
 && \mu_{q}=\sqrt{m_{q}^2 + k_{q}^2}  + \left(\frac {g_v} {m_V}\right)^2 n_Q \nonumber \\ 
   &&   \mu_{l}=\sqrt{m_{l}^2 + k_{l}^2} . 
      \label{equi}
 \end{eqnarray}
where \( k_{q} \) denotes the Fermi momentum of the quarks, while \( k_{l} \) represents the Fermi momenta of electrons and muons.
 
Then we can use the local charge conservation condition 

\begin{equation}
    \frac{2}{3}\rho_u - \frac{1}{2}(\rho_d + \rho_s) - \rho_e - \rho_\mu = 0
\end{equation}
where $\rho_{i}$ represents the number density of each particle species. In addition, the beta equilibrium condition is imposed through the relations
\begin{equation}
     \mu_s = \mu_d = \mu_u + \mu_e, \quad 
\mu_\mu = \mu_e.
\end{equation}
respectively.

The above mentioned conditions guarantee that weak interaction processes remain in equilibrium while maintaining the overall electrical neutrality of strange quark matter. Also, we must emphasize that beta equilibrium and local electric charge conservation must be maintained even under conditions of strange matter instability - a necessary condition for the stability of the 
star. 

For a stable hybrid star with a quark core, the vMIT model parameters $\{B^{1/4}, G_{v}, b_{4}\}$ must ensure that strange quark matter remains {\textit{energetically disfavored}} 
\cite{Lopes:2020btp, Lopes2021c}. In other words, the previous parameters must lie outside of the stability window.  Otherwise, the entire hadronic layer would convert to deconfined matter, resulting in a strange star rather than a hybrid star \cite{Graeff_2019,kau17}.

\begin{table*}[!htb]
\caption{Nuclear matter properties for the models discussed in the text. The binding energy $E/A$, compressibility modulus $K_0$, symmetry energy $S_0$, and its slope $L_0$ are given in MeV, and the effective mass parameter ${M^*/M}$ is dimensionless. All quantities are calculated at the saturation density $n_0$ (in fm$^{-3}$).}
\label{tab:propriedades}
\begin{tabular}{l|cccccc}
    \hline\hline
    Model & {$n_0$} & ${M^*/M}$ & ${E/A}$ & $K_0$ & $S_0$ & $L_0$ \\ \hline
    {NL3$^*\omega \rho$} & 0.150 & 0.59& 16.3 & 258  & 30.7 & 42  \\
    {EL3$\omega \rho$} & 0.156& 0.69 & 16.2 & 256  & 32.1 & 66 \\ \hline \hline
\end{tabular}

\end{table*}

\subsection{Hadron-Quark Phase Transition}

There is a well-documented tension between LQCD and effective models regarding the nature of the QCD phase transition. LQCD predicts a smooth crossover transition at low chemical potential, occurring at a temperature of approximately 160--170 MeV \cite{aoki2006order, bellwied2015qcd, luostudy}. In contrast, effective models indicate a first-order phase transition at high densities. Between these regimes, the transition is expected to culminate in a critical end point (CEP), where the phase transition becomes second order \cite{Biesdorf:2023icx}.
However, the existence and exact location of the CEP remain uncertain \cite{bazavov2017skewness, bazavov2017qcd}.

We assume that the hadron-quark deconfinement transition is a first-order phase transition, as predicted by effective models in the high-density region of the QCD phase diagram. It is important to emphasize that the exact properties of the phase transition are model-dependent. The characteristics of the transition depend on the quark and hadron EOS models used. For instance, \cite{Fukushima_2011} suggests that, within the Polyakov loop formalism, the onset of the phase transition at zero temperature requires a chemical potential greater than $\mu = 1050$ MeV.

The thermodynamic description of this process involves matching the EOS of the two phases, and identifying the point of phase coexistence. A first-order phase transition can 
be obtained in two ways: according to either the Maxwell or Gibbs constructions. In Maxwell construction, the two phases are distinct and maintain local charge conservation. In contrast, the Gibbs construction allows quarks and hadrons to coexist up to a certain baryonic density, with global charge conservation applied.
To determine which phase construction to apply, the surface tension serves as the primary criterion for describing the phase transition appropriately. If the surface tension exceeds 60 MeV/fm$^2$, the Maxwell construction is adequate \cite{Voskresensky:2002hu, Maruyama:2007ey}. Conversely, if it is lower, the Gibbs construction must be preferred. In this study, we choose the Maxwell construction due to the uncertainties surrounding the value of the surface tension \cite{Pinto:2012aq, Lugones:2013ema, Lugones:2016ytl, Lugones:2018qgu} and the 
fact that the choice of either Gibbs or Maxwell constructions has a minimal impact on the mass-radius relations of neutron stars.

The thermodynamic description of the Maxwell construction involves matching the EOS for the two phases to determine the phase coexistence point -- where $P_{\rm hadron}=P_{\rm quark}=P_0$ (mechanical equilibrium) and $\mu_{\rm hadron}=\mu_{\rm quark}=\mu_0$ (chemical equilibrium). The location of the coexistence point is determined from the equations of state of both the hadronic and the
quark phases, so it is expected that the choice of the vector MIT bag model free parameters significantly influences the transition point for a given baryonic composition of the hadronic phase. 
One can observe that the parameters of hadronic equations of state (EOSs) are fixed. Hence, only the vector MIT bag model exhibits variations in its slope ($P \times \mu$ plane).

\section{ Macrophysics}
\label{tov_tidal}

In this section, we investigate the imprint of different phase transitions on the structure and tidal deformability of hybrid stars. First we present the stellar structure differential equations for an equilibrium compact star. With these equations at hand, it is possible to obtain the mass and radius. Second, we present the differential equation of tidal deformability, whose results can be compared with recent observations of gravitational waves.

\subsection{Stellar structure}

The hydrostatic equilibrium configuration of a nonrotating compact star is determined by the Tolman-Oppenheimer-Volkoff (TOV) equations of relativistic stellar structure \cite{TOV}. In order to obtain those equations, we have to select the following static spherically symmetric metric,
\begin{equation}\label{metric}
    ds^2 = - e^{\nu}dt^2 + e^{\lambda} dr^2 + r^2(d\theta^2 + sin^2 \theta d \phi^2),
\end{equation}
in addition, we consider the energy-momentum tensor of a perfect fluid, as given below
\begin{equation}\label{tem}
    T_{\mu\nu}=p g_{\mu\nu}+(\epsilon+p)u_{\mu}u_{\nu},
\end{equation}
where $\epsilon$, $p$, and $u_{\mu}$ are, respectively, the energy density, pressure, and four-velocity of a fluid element. 

Once we have the metric (\ref{metric}) and the energy-momentum tensor (\ref{tem}), we proceed to use these quantities as input for the Einstein field equations(\ref{einstein})
\begin{equation}
 R_{\mu\nu} - \frac{1}{2} g_{\mu\nu} R =8\pi T_{\mu\nu}\label{einstein}.
\end{equation}

After some algebraic work, we can obtain the TOV equations that describe the hydrostatic equilibrium. These equations reads as
\begin{align}
\label{tov1}
\frac{dp}{dr} ={}& - \frac{\epsilon m}{r^2}\bigg(1 + \frac{p}{\epsilon}\bigg)
	\bigg(1 + \frac{4\pi p r^3}{m}\bigg)\left(1 - \frac{2m}{r}\right)^{-1},  \\
\label{tov2}
\frac{d \nu}{dr} ={}& - \frac{2}{\epsilon} \frac{dp}{dr} \bigg(1 + \frac{p}{\epsilon}\bigg)^{-1}, \\
\label{tov3}
\frac{dm}{dr}={}& 4 \pi r^2 \epsilon, 
\end{align}
where $p$ is the pressure, $\epsilon$ is the energy density, $m$ is the gravitational mass and $\nu$ is the metric potential. The local pressure,  energy density and baryonic number density are related by the equation of state previously discussed in Sec. II.

In order to solve the TOV equations, we start the integration at the star center using the following initial conditions $m(0) =0$  and $p(0) = p_c$, where $p_c$ is the central pressure. The integration is performed until the pressure becomes null and at this point we define the radius of the star. Also considering continuity of the gravitational field, we impose the following matching condition for the metric: $\nu(R)= \ln ( 1- {2M}/{R} )$.

\subsection{Tidal Deformability}

In a binary system, the shape of each star becomes deformed due to the external tidal field ($\phi_{ij}$) exerted by its companion. As a result, the stars develop a quadrupole moment ($Q_{ij}$), which is linearly proportional to the tidal field, with the proportionality constant represented by $\lambda$. This relationship is expressed as:
\begin{equation}
    Q_{ij} = -\lambda \phi_{ij}.
\end{equation}
The dimensionless tidal deformability ($\Lambda$) is connected to the compactness parameter, defined as $C = M/R$, through the relation:

\begin{equation}
    \Lambda \equiv \frac{\lambda}{M^{5}} = \frac{2 k_{2}}{3 C^{5}}. \label{stidal}
\end{equation}

The tidal deformability parameter quantifies how easily a compact object---in this case, a hybrid star in a binary system (HS-HS merger)---is deformed under the influence of an external tidal field ($\phi_{ij}$). A larger $\lambda$ value indicates a more deformable compact object.

 Here we provide a summary of the calculation procedure of the dimensionless tidal parameter $\Lambda$.
At first one must integrate the following first order differential equation \cite{PhysRevD.80.084035,PhysRevD.81.123016,PhysRevD.109.083021}, 
\begin{equation}
 r\frac{dy}{dr} +y^2 + yF(r) +r^2Q(r) = 0, \label{EL15}
\end{equation}
where the coefficients $F(r)$ and $Q(r)$ are given by
\begin{align} 
F(r) ={}& {\left[1-4 \pi r^{2}(\epsilon-p)\right]\left[1-\frac{2 m}{r}\right]^{-1} } , \\ 
Q(r) ={}& 4 \pi\left[5 \epsilon+9 p+\frac{\epsilon+p}{c_{s}^{2}}-\frac{6}{4 \pi r^{2}}\right]\left[1-\frac{2 m}{r}\right]^{-1}   \nonumber \\ 
& -\frac{4 m^{2}}{r^{4}}\left[1+\frac{4 \pi r^{3} p}{m}\right]^{2}\left[1-\frac{2 m}{r}\right]^{-2} ,
\label{EL17}
\end{align}
with $\epsilon$, $p$, $m$, and  $c_{s}^{2} = d p / d \epsilon$ being respectively the energy density, pressure, mass, and the squared speed of sound.

To integrate the differential equation \eqref{EL15}, we start at the center of the star with the initial condition $y(0) = 2$, using the mass, density, pressure, and speed of sound defined at each point inside the star (these quantities are obtained from the solution of the TOV equations). When the surface of the star is reached we stop the integration and finally get $y_R = y(r = R)$. This last quantity is used to obtain the second-order Love number, defined by
\begin{align} 
k_{2}= {}& \frac{8 C^{5}}{5}(1-2 C)^{2}\big[2-y_{R}+2 C\left(y_{R}-1\right)\big]\nonumber \\ & \times\big\{2 C\left[6-3 y_{R}+3 C\left(5 y_{R}-8\right)\right]  \nonumber \\ & +4 C^{3}\left[13-11 y_{R}+C\left(3 y_{R}-2\right)+2 C^{2}\left(1+y_{R}\right)\right]  \nonumber \\  & +3(1-2 C)^{2}\left[2-y_{R}+2 C\left(y_{R}-1\right)\right] \ln (1-2 C)\big\}^{-1},
\end{align}
Once $k_{2}$ is defined, it is possible to calculate $\Lambda$ by the use of equation (\ref{stidal}).

Furthermore, in order to calculate the dimensionless tidal deformability of each hybrid star in the inspiral, we use the chirp mass. As two compact objects orbit one another, they lose orbital energy through gravitational-wave emission. This dissipation causes them to inspiral, drawing closer and accelerating. The gravitational-wave frequency therefore increases with time, generating a characteristic “chirp” signal whose pitch rises rapidly. The key parameter governing this frequency evolution is the chirp mass, given by
\begin{equation}
\mathcal{M} = \frac{(m_1 m_2)^{3/5}}{(m_1 + m_2)^{1/5}}.
\end{equation}
The values  in \footnote{The properties of gravitational-wave sources are determined by comparing observational data with a post-Newtonian waveform model in the frequency domain. The collaboration\cite{LIGOScientific:2017vwq} performs a Bayesian analysis across a frequency range of 30--2048 Hz.} refer to the low-spin prior $(|\chi| \leq 0.05)$, as both $m_{1}$ and $m_{2}$ are consistent with the masses of all known binary neutron star systems\cite{LIGOScientific:2017vwq} with 90$\%$ credible intervals.

The chirp mass $\mathcal{M} = 1.188^{+0.004}_{-0.002} \ M_{\odot}$, with the primary mass $m_{1} \in [1.36, 1.60] \ M_{\odot}$, the secondary mass $m_{2} \in [1.17, 1.36] \ M_{\odot}$, and the total mass $m_{\text{tot}} = 2.74^{+0.04}_{-0.01} \ M_{\odot}$. The combined dimensionless tidal deformability - to  leading order in $\Lambda_{1}$ and  $\Lambda_{2}$ is 

\begin{equation}  
\tilde{\Lambda} = \frac{16}{13} \frac{(m_1 + 12 m_2) m_1^4 \Lambda_1 + (m_2 + 12 m_1) m_2^4 \Lambda_2}{(m_1 + m_2)^5},
\end{equation}
with $\tilde{\Lambda}$ - assuming a uniform prior -  and $\Lambda(1.4) \leq 800$ for low-spin priors. 
The probability density shown in the plots of $\Lambda_{1}$ against $\Lambda_{2}$ should be considered as a conservative upper limit, as the post-Newtonian model tends to overestimate these values\cite{LIGOScientific:2017vwq}.

\subsection{Approximate Universal Relations}
Approximate universal relations (i.e. insensitive to the equation of state) establish a direct link between the microscopic and macroscopic properties of neutron stars, providing valuable understanding of nuclear matter. One way to break the degeneracies in neutron star properties is by using these relations. For instance, the authors of \cite{Yagi_2016} found the binary Love relation. Using individual Love numbers, they constructed a dimensionless symmetric combination 
\[
\bar{\lambda}_s = \frac{\bar{\lambda}_1 + \bar{\lambda}_2}{2},
\] 
and a dimensionless antisymmetric combination, 
\[
\bar{\lambda}_a = \frac{\bar{\lambda}_1 - \bar{\lambda}_2}{2},
\] 
where \(\bar{\lambda}_A = \lambda_A / m_A^5\), and the index A=1,2. 
The relation between \(\bar{\lambda}_s\) and \(\bar{\lambda}_a\) holds for \(q = m_2/ m_1 < 1\) and \(m_1 \in (1 M_{\odot}, m_{1}^{\text{max}})\), where \(m_{1}^{\text{max}}\) is the maximum mass of a stable neutron star for a given EOS. The authors verify this relation using the following realistic EOSs: WFF1\cite{Wiringa:1988tp}, MS1\cite{Mueller:1996pm}, LS220\cite{LS}, and AP4\cite{APR}. This relation enables the inference of individual tidal deformabilities, which are currently not directly measurable from gravitational wave data.

Another relation, such as the Love-C relation \cite{science.1236462} between tidal deformability and neutron star compactness, can be used to estimate the neutron star radius. Also the relation which links the moment of inertia, tidal deformability, and quadrupole moment ("I-Love-Q"), is remarkably precise, holding with approximately percentage-level accuracy\cite{science.1236462}.
Furthermore, we utilize an approximately universal relation proposed by the authors \cite{Saes:2021fzr}, which connects $\Upsilon = p_{c}/\epsilon_{c}$
 with \(C / \bar{I} / \Lambda\), where \(C\) represents the compactness of the star, 
\(\bar{I}\) denotes the dimensionless moment of inertia, and \(\Lambda\) is the dimensionless tidal deformability. 
\(\Upsilon\) can be interpreted as a mean of the stiffness of nuclear matter within the star. It depends on microscopic properties and is defined as the ratio of central pressure to central energy density.
We compare our results with the polynomial fit, which incorporates 24 realistic hadronic EOSs and one for a hybrid star with a quark core, QHC19; it is found that the fit is a fifth-degree polynomial. It should be mentioned that this fit is only valid for \( C > 0.05 \).
We observe that our results fall within the error band of this fit, indicating agreement with these relations. In this way, it is even possible to determine the values of the dimensionless moment of inertia from \(\Upsilon\) with a good approximation.

We also have compared the inferred median results for the cases with two parameters $(G_v, B^{1/4})$ with the approximate universal relation $C-\Lambda$ found by \cite{YAGI20171}. The fitted relation uses the following EOS.
For neutron stars consisting exclusively of conventional (baryonic) matter, the selected set of equations of state are  APR, AP3, AP4, SLy, SGI, SV, SkI4, LS220, Shen, ENG, MPA1, MS1, MS1b, WFF1, WFF2, DBHF(2) (A), GA-FSU2.1, G4, GCR, and MPa.
In contrast, for neutron stars that include exotic constituents, they adopt equations of state such as ALF5, SGI-YBZ6-S3, SkI4-YBZ6-S3, NlY5KK, GA-FSU2.1-180, H4, and MPaH. This latter set encompasses models featuring kaon condensation, hyperonic matter, and quark--gluon plasma phases.
The relation is given by, 
\begin{equation}
C = a_0 + a_1 \ln \Lambda + a_2 (\ln \Lambda)^2,
\end{equation}
where $C = M/R$ is the compactness and the coefficients were fit to be 
$a_0 = 0.360$, $a_1 = -0.0355$, and $a_2 = 0.000705$. The relation is accurate to within $6.5\%$.

\section{Bayesian Inference Methodology}

Bayesian inference is a robust statistical approach for estimating model parameters from data \cite{Malik:2022zol,Imam:2021dbe,Wesolowski:2015fqa,Furnstahl:2015rha,Ashton:2018jfp,Huang:2023grj,Char:2023fue,Imam:2024gfh}. By applying Bayes's theorem \cite{Bayes1763}, prior knowledge about the parameters is systematically updated in light of new data, resulting in posterior distributions that reflect the refined understanding of these parameters. In this section, we present Bayesian analysis as a tool for constraining the vMIT model parameters and, consequently, the onset of deconfined quark matter.

The main idea is to construct the posterior probability distribution of the model parameters via Bayes's theorem, which reads as
\begin{equation}
P(\theta|D, M) = \frac{P(D|\theta, M) P(\theta, M)}{P(D, M)}.
\end{equation}
This approach allows for the calculation of the probability distribution of a set of parameters \(\theta\), given the data \(D\) within a specific model \(M\). The resulting distribution, \(P(\theta \mid D, M)\), is known as the posterior probability. It is derived by combining prior knowledge, represented by the prior probability \(P(\theta, M)\), which reflects the understanding before accounting for the data, with the likelihood function \(P(D \mid \theta, M)\), which represents the statistical information derived from probabilistic models. The probability of the observed data \(D\), known as the evidence \(P(D, M)\), serves as a normalization factor since it does not depend on the parameters.

\begin{table}[!]
\centering
\begin{tabular}{c c c c c}
\hline
Stars & Radius [km]  & Mass $[M_{\odot}]$ & CI [$\%$] & $\rho$ \\
\hline
PSR J0030+0451 (PDT-U)\cite{Vinciguerra:2023qxq} & $13.12_{-1.21}^{+1.35}$  & $1.41_{-0.19}^{+02}$ &     68 &  0.95 \\
PSR J0740+6620  (ST-U) \cite{Salmi:2024aum}  & $12.38_{-0.78}^{+1.03}$  & $2.072_{-0.066}^{+0.067}$&   68 & 0.0\\
GW 170817\cite{Abbott2018_GW170817_Radii} (m1) & $10.08_{-1.7}^{+2.0}$ & $1.49_{-0.1}^{+0.1}$     &  90  & 0.0 \\
GW 170817\cite{Abbott2018_GW170817_Radii} (m2) & $10.07_{-1.5}^{+2.1}$ & $1.255_{-0.1}^{+0.1}$      &  90 & 0.0\\
\hline
\end{tabular}
\caption{The mass and radius of compact stars are employed as constraints in the likelihood function. The $\sigma$ values are defined as the mean of the upper and lower bounds. CI denotes the credible interval, and $\rho$ represents the correlation parameter. }
\label{tab1}
\end{table}

For the inference performed in this study we have the  vMIT model parameters being $\theta=\{G_{v},B^{1/4},b_{4}\}$ 
and assume an uniform prior distribution within the ranges $B^{1/4} \in [150,180 ]$ MeV, $G_{v} \in [0.2 , 0.3]$ fm$^{2}$ and $b_{4} \in [0, 2]$. The data set $D$ are the masses and radii, as presented in Table \ref{tab1}. The full likelihood function is composed of contributions from the phase transition, mass-–radius constraints, and the minimum allowed mass. \\

{\bf{Likelihood Function for the Phase Transition}}

We follow the approach of Ref.~\cite{Albino:2024ymc,Albino2025puc} to constrain the density range of the phase transition. We assume a super-centered Gaussian at $n_b = 0.275~\mathrm{fm}^{-3}$, with $p = 5$ and standard deviation $\sigma = 0.08~\mathrm{fm}^{-3}$. These parameters imply that $0.195 \leq n_{b,{\mathrm{trans}}} \leq  0.355~\mathrm{fm}^{-3}$.
\begin{equation}
\log(\mathcal{L}_{\mathrm{phT}}) =
- \left[\frac{(n_b,_{\mathrm{trans}} - 0.275)^2}{2(0.08)^2}\right]^5 
\end{equation}
where the selected values for the super-centered Gaussian are intended to guarantee a large quark core.

{\bf{Likelihood Function for the mass-radius}}

We adopt a bivariate Gaussian distribution for the mass--radius measurements presented in Table~\ref{tab1}.

\begin{align}
& \log\!\left(\mathcal{L}(\mathrm{I})\right)
=
\log\!\left(
\frac{1}{2\pi\,\sigma_m\,\sigma_r\,\sqrt{1-\rho^2}}
\right) \\ \nonumber
&
-\frac{1}{2(1-\rho^2)}
\Bigg[
\left(\frac{m-\mu_m}{\sigma_m}\right)^2
+
\left(\frac{r-\mu_r}{\sigma_r}\right)^2
 \\ \nonumber & -2\rho
\left(\frac{m-\mu_m}{\sigma_m}\right)
\left(\frac{r-\mu_r}{\sigma_r}\right)
\Bigg].
\end{align}
Where I=NICER, GW.
Note that when 
$\rho=0$, the expression reduces to the sum of two Gaussian distributions. The same likelihood function is used for both NICER and GW, but with different correlation parameters ($\rho$). The subscripts \textit{m} and \textit{r}  refer to the values (observed) listed in Table~\ref{tab1}. On the other hand, \textit{m} and \textit{r} correspond to a particular set of parameters in the vMIT model and represent the theoretical values.

{\bf{Likelihood Function for the Minimum Mass Constraint}}

We restrict our inference to the minimum mass of PSR J0740+6620, adopting a lower bound of $M \ge 2.006,M_{\odot}$.
\begin{equation}
\log\left(\mathcal{L}_{\mathrm{max\,mass}}\right) =
-\infty, \quad \text{for } \quad M_{\max}  < 2.006~M_{\odot} .
\end{equation}

The full likelihood function is given by
\begin{equation}
\begin{aligned}
\mathcal{L}_{\mathrm{FULL}} =\;& 
\mathcal{L}_{\mathrm{phT}}
 \mathcal{L}(\mathrm{PSR\ J0030+0451}) \\
&\mathcal{L}(\mathrm{PSR\ J0740+6620}) 
\mathcal{L}(\mathrm{GW\ 170817\ (m1)})\\
&\mathcal{L}(\mathrm{GW\ 170817\ (m2)}) 
\mathcal{L}_{\mathrm{max\,mass}}.
\end{aligned}
\end{equation}

In this way, \( P(D \mid \theta, M) = \mathcal{L}_{\mathrm{FULL}} \). Furthermore, to improve stability in numerical code, we use the logarithmic scale in Bayes's theorem.
To compute the evidence, we used the \texttt{dynesty} sampler \cite{dynesty} with \texttt{nlive = 200}, which performs this calculation automatically.
 The \texttt{corner}  Python package\cite{ForemanMackey2016} is utilized to create contour plots, while other plots are generated using Matplotlib\cite{matplotlib}.

\section{Results and Discussion} \label{sec_res}

In this section, we analyze the results from the Bayesian inference. As it is well known, hydrid stars are composed of a core of deconfined quark matter and a hadronic envelope composed of a hadronic matter phase.  Therefore, it is necessary to consider a suitable equation of state for each phase in the hybrid star. For this objective, for the case of quark matter we have use the vMIT parameter set, for the case of hadronic matter without hyperons we use the parameter sets \( NL3^{*}\omega\rho \), \( EL3\omega\rho \) and, for the case of hadronic matter with hyperons we use \( EL3\omega\rho Y \). It is worth mentioning that the parameter set \( NL3^{*}\omega\rho \) results in a stiff EOS, while \( EL3\omega\rho \) leads to an intermediate EOS. In this way, by applying the Maxwell construction and Bayesian inference with the aforementioned priors discussed in the previous section and using the full likelihood function, we aim to determine the best vMIT parameter set for both the two-parameter case $(G_v, B^{1/4})$ and the three-parameter case $(G_v, B^{1/4}, b_4)$, which most accurately describe the NICER observations and the inferred masses from the GW170817 event.

The following results correspond to the EL3$\omega\rho$ set for two parameters.

\begin{figure}[H]
     \centering
     \includegraphics[scale=0.5]{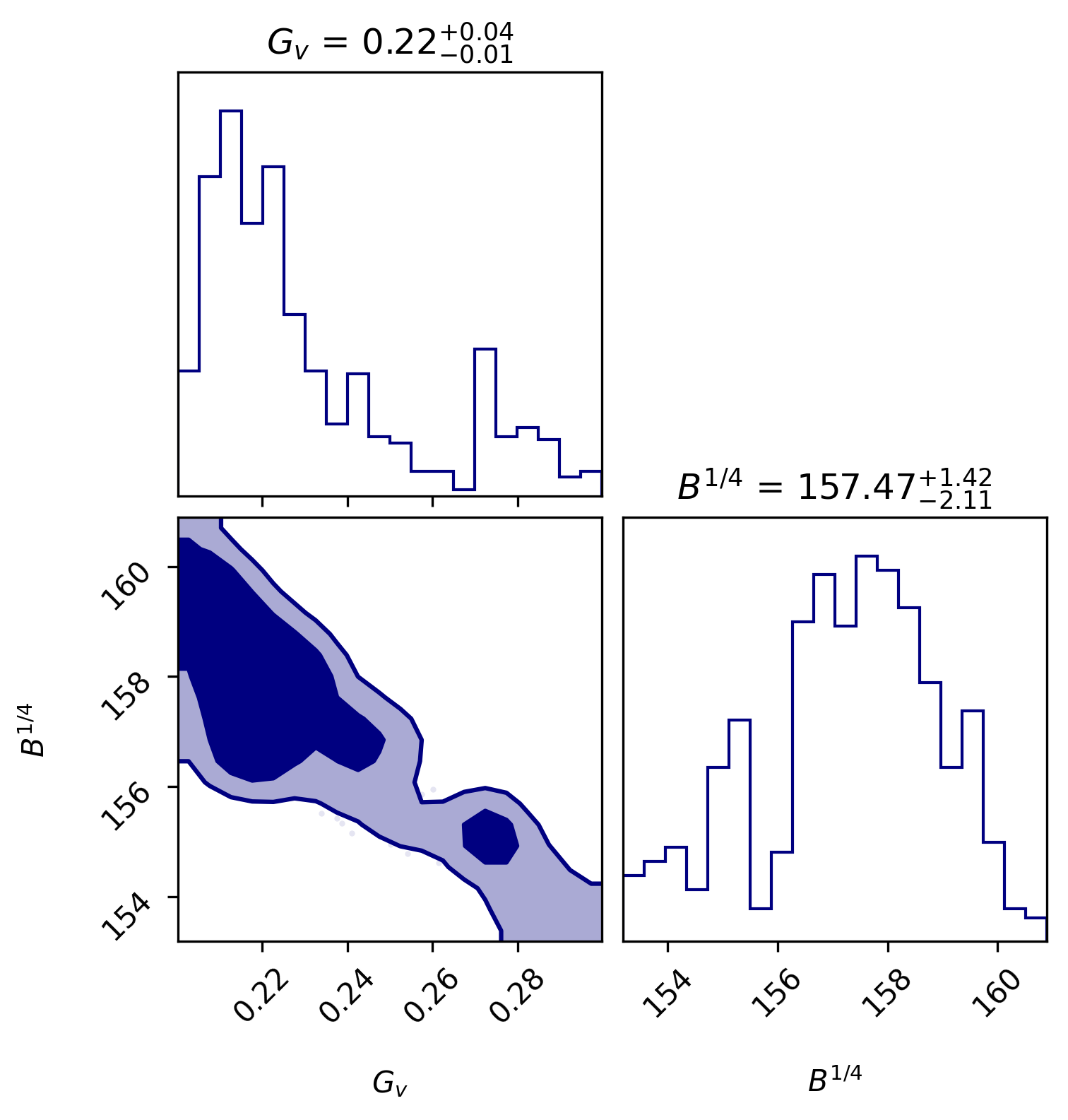}
        \caption{The results for the inference. For quark matter we have used the two vMIT parameters with (0.22, 157.47). For hadronic matter we use the EL3$\omega\rho$ parametrization.}
        \label{fig1a}
 \end{figure}

 \begin{figure}[H]
        \hspace{5.7cm}
        \includegraphics[scale=0.5]{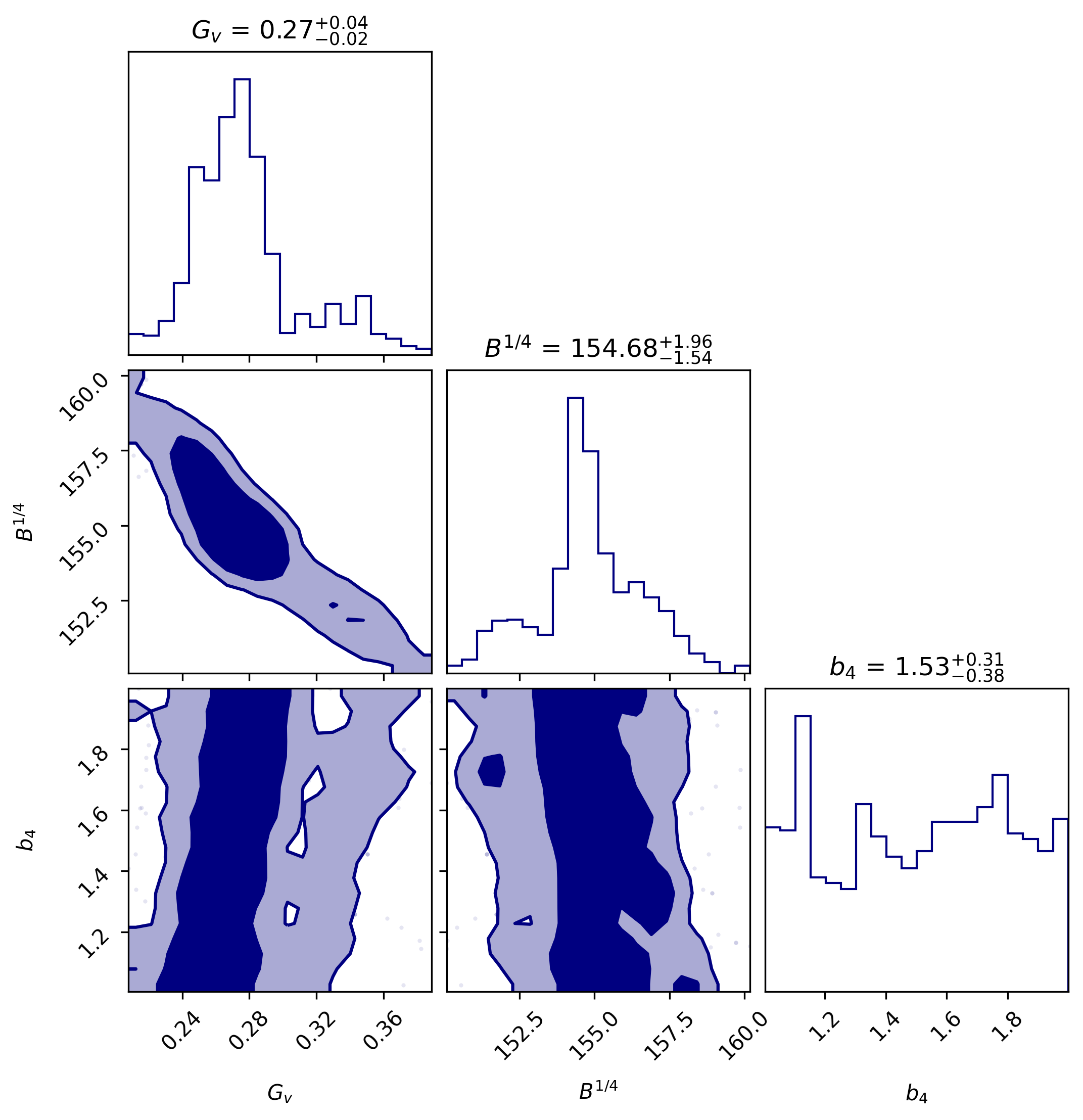}
        \caption{The results for the inference. For quark matter we have used the three vMIT parameters with (0.27, 154.68,1.53). For hadronic matter we use the EL3$\omega\rho$ parametrization}
        \label{fig1b}
    \end{figure}
The dark-to-light blue contours in Figs. \ref{fig1a} and \ref{fig1b} represent the 68$\% $ credible region and  $95\%$ credible region, respectively.

\begin{figure}[H]
    \centering
    \begin{subfigure}[t]{0.47\textwidth}
        \centering
        \includegraphics[scale=0.42]{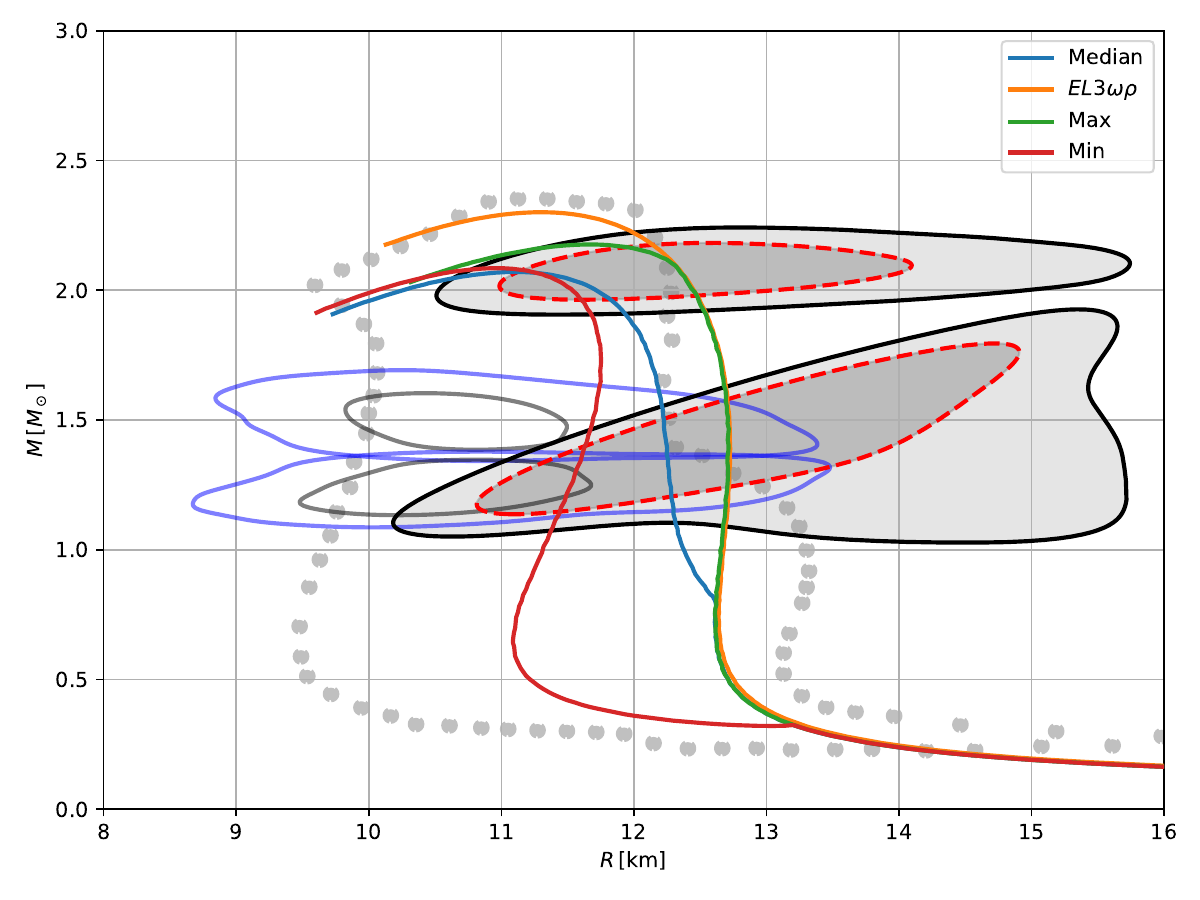}
        \caption{}
        \label{fig3a}
    \end{subfigure}
    \hfill
    \begin{subfigure}[t]{0.47\textwidth}
        \centering
        \includegraphics[scale=0.47]{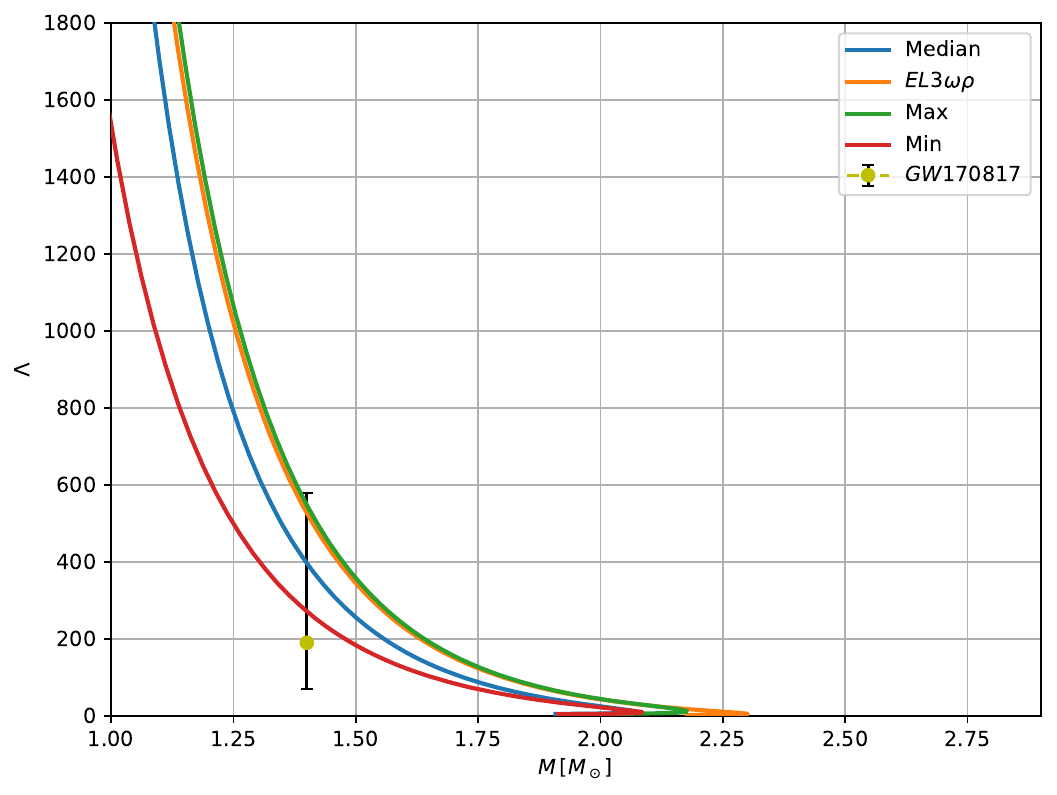}
        \caption{}
        \label{fig3b}
    \end{subfigure}

    \vspace{0.25cm}

    \begin{subfigure}[t]{0.49\textwidth}
        \centering
        \includegraphics[scale=0.57]{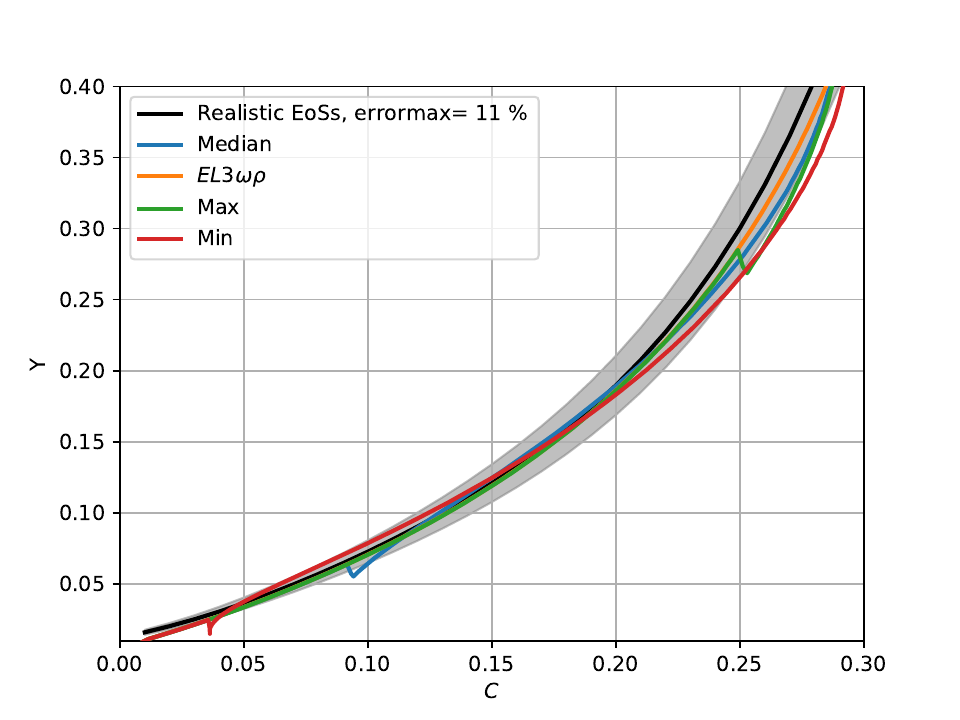}
        \caption{}
        \label{fig3c}
    \end{subfigure}
    \hfill
    \begin{subfigure}[t]{0.49\textwidth}
        \centering
        \includegraphics[scale=0.57]{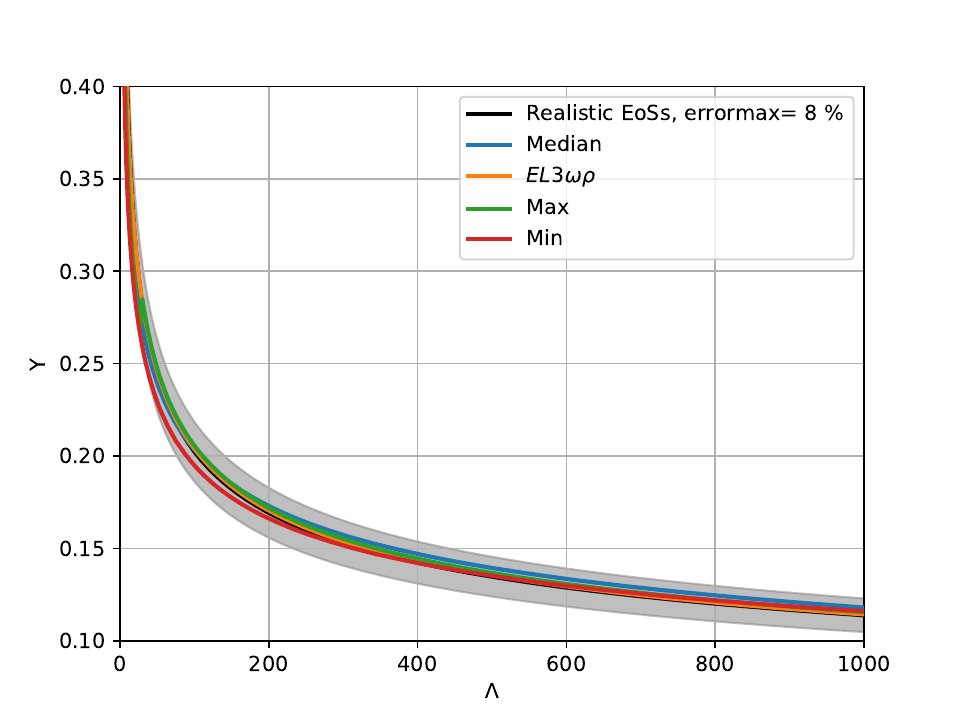}
        \caption{}
        \label{fig3d}
    \end{subfigure}

    \caption{ (a) Mass--radius diagram obtained from the vMIT inference for the parameter set $EL3\omega\rho$. The transition density corresponding to the lower bound (min) lies below the imposed numerical cutoff ($> 1.5\,n_0$) and $\mathcal{L}_{\mathrm{phT}}$. The median values (0.22, 157.47), represented by the blue solid line, lie within the pQCD constraints and yield a transition mass of $\simeq 0.8\,M_{\odot}$. Furthermore, the green solid line (max) results in a small quark core and lies outside the pQCD constraints. The natural log evidence for this model is $-4.54$. The minimum (min) and maximum (max) solid lines correspond to the lower and upper bounds of the 68\% credible interval for the mass--radius relation. (b) Dimensionless tidal deformability $\Lambda$ as a function of the stellar mass. The median value (blue solid line) lies close to $\Lambda_{1.4} = 400$, consistent with the pQCD constraint shown in panel (a). (c) The lower-bound solution (red solid line) is disfavored in the $\Upsilon$--$C$ relation, where $C$ denotes the stellar compactness \cite{Saes:2021fzr}. The remaining inferred results lie within the error band derived from the set of 24 realistic EOSs considered in Ref.~\cite{Saes:2021fzr}, except for the discontinuity associated with the first-order phase transition (cusp). This feature arises because the central pressure remains constant across the transition while the central energy density changes abruptly, causing a corresponding jump in $\Upsilon = P_c/\epsilon_c$. We should point out that, before the cusp, only neutron-star configurations are present, whereas beyond the cusp, hybrid-star configurations appear. (d) $\Upsilon$ as a function of $\Lambda$ \cite{Saes:2021fzr}. All inferred results fall within the error band constructed from the 24 realistic EOSs analyzed in Ref.~\cite{Saes:2021fzr}.}

\end{figure}

In Fig. \ref{fig3a}, only the median inferred value lies within the gray dots pQCD constrained EOS\cite{PhysRevLett.120.172703} ($\Lambda$ < 400) and has $n_{b,trans}>1.5 n0$. 
These results  are in good agreement with the observed data of mass and radius of compact stars as shown in Table \ref{tab1}. The area indicated by the gray dots is generated by considering the EOS given by Chiral Effective Field Theory ($\chi EFT$) for densities up to $n = 1.1n_0$, since its uncertainties at this density reach $\pm 24\%$. The advantage of $\chi EFT$ lies in its ability to estimate errors for its predictions at each loop calculation and explain the hierarchy of two, three, and weaker higher-body forces~\cite{Hebeler:2009iv}. They employed the "hard" or "soft" EOS from Ref.~\cite{Hebeler:2009iv}, which correspond to the most extreme EOS  permitted at low densities. For the intermediate densities, the politropic form $p_{i}=k_{i}n^{\gamma_{i}}$ 
was used with three or four polytropes up to $\mu_{B}=2.6~GeV$—with an uncertainty level of $\pm 24 \%$. Both $k$ and $\gamma ~ (\text{adiabatic index})$ are constants determined for each range density (index i) of the EOS. This chemical baryon potential corresponds to the deconfined quark matter at high density ($40n_{0}$) obtained from the calculation of NNLO pQCD\cite{PhysRevD.81.105021}. Thus, the gray dots correspond to the previous EOS for four polytropes, which can describe both the GW170817 event and the two solar mass constraints. 

In Fig.~\ref{fig3b}, although the dimensionless tidal deformability was not included in the likelihood function, all results are consistent with the observational estimate of $\Lambda_{1.4} = 190^{+390}_{-120}$ \cite{LIGOScientific:2017vwq}.

In Figs. \ref{fig3c} and \ref{fig3d}, we can note that, except for the phase transition in Fig.3 (c), it is possible to estimate the dimensionless moment of inertia from $\Upsilon$, as these lie within the error band.
     
In Fig \ref{fig4} the results corresponding to the parameter set $EL3\omega\rho$ for three parameters (0.27, 154.68,1.53) are displayed.

\begin{figure}[H]
    \centering

    \begin{subfigure}[t]{0.47\textwidth}
        \centering
        \includegraphics[scale=0.42]{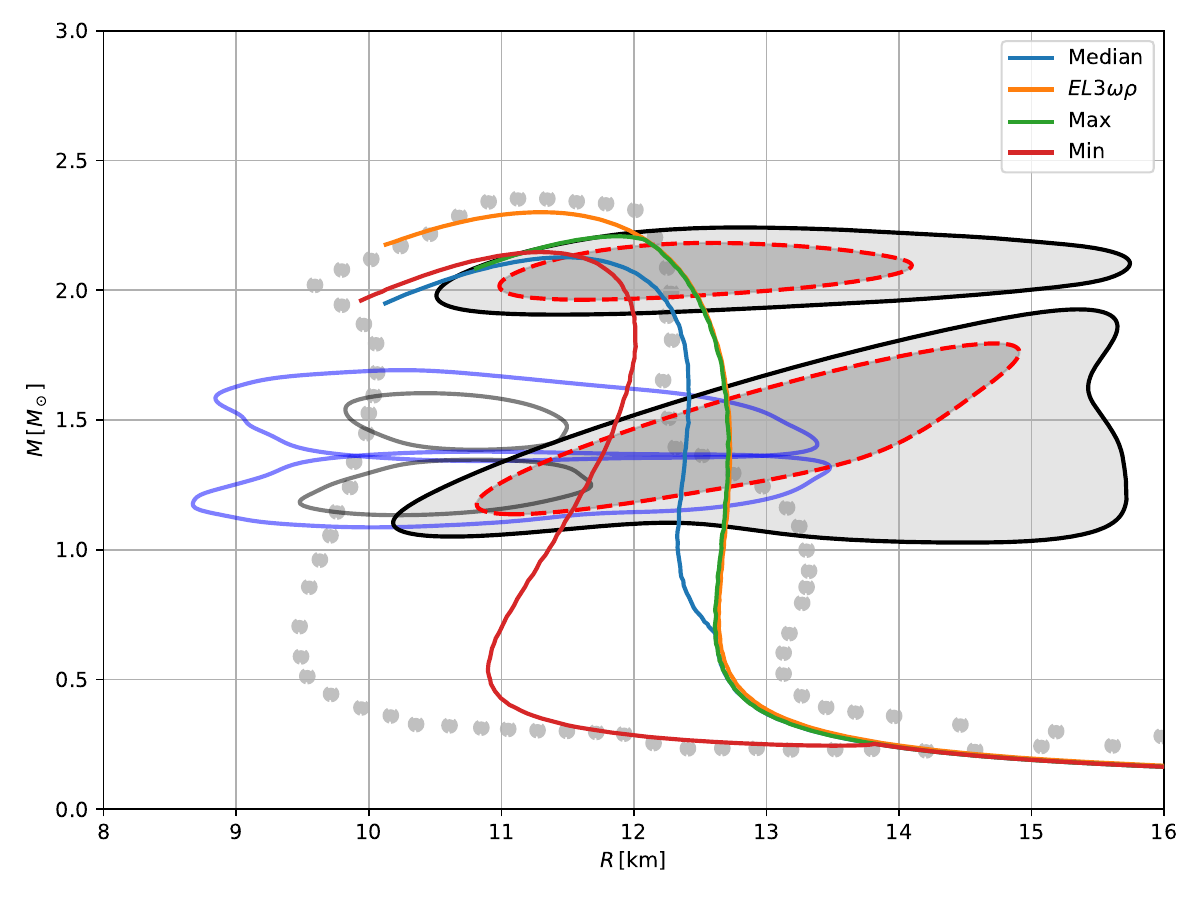}
        \caption{}
        \label{fig4a}
    \end{subfigure}
    \hfill
    \begin{subfigure}[t]{0.47\textwidth}
        \centering
        \includegraphics[scale=0.47]{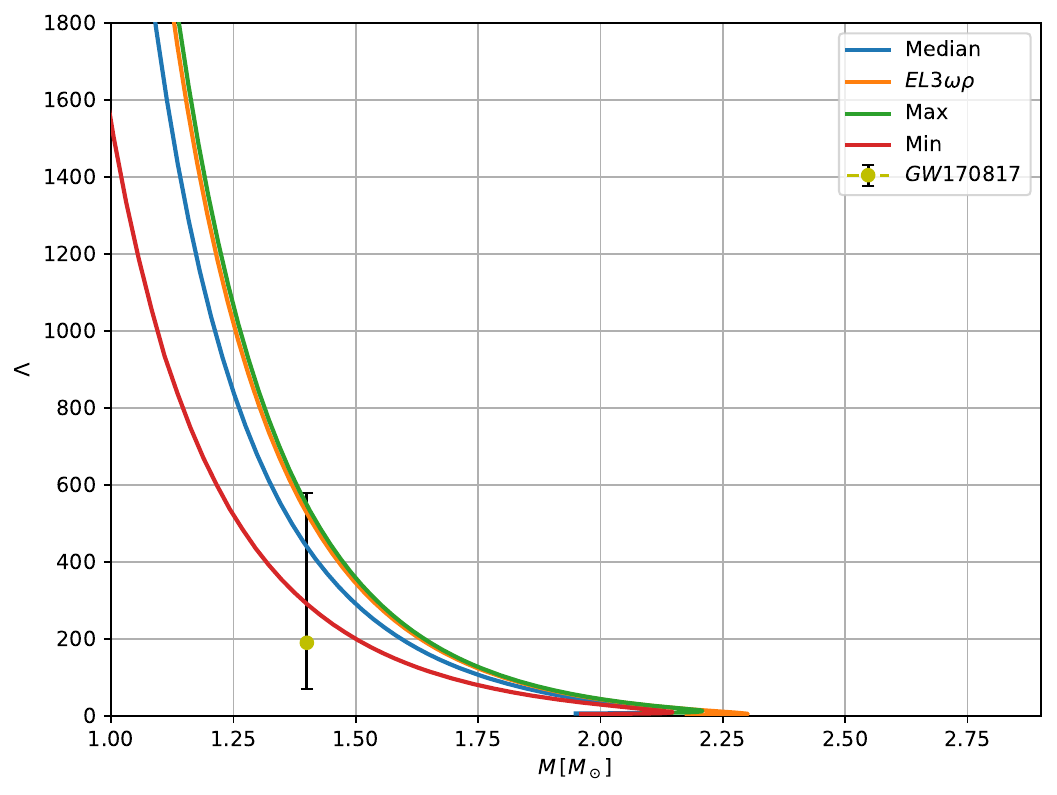}
        \caption{}
        \label{fig4b}
    \end{subfigure}

    \vspace{0.5cm}

    \begin{subfigure}[t]{0.49\textwidth}
        \centering
        \includegraphics[scale=0.57]{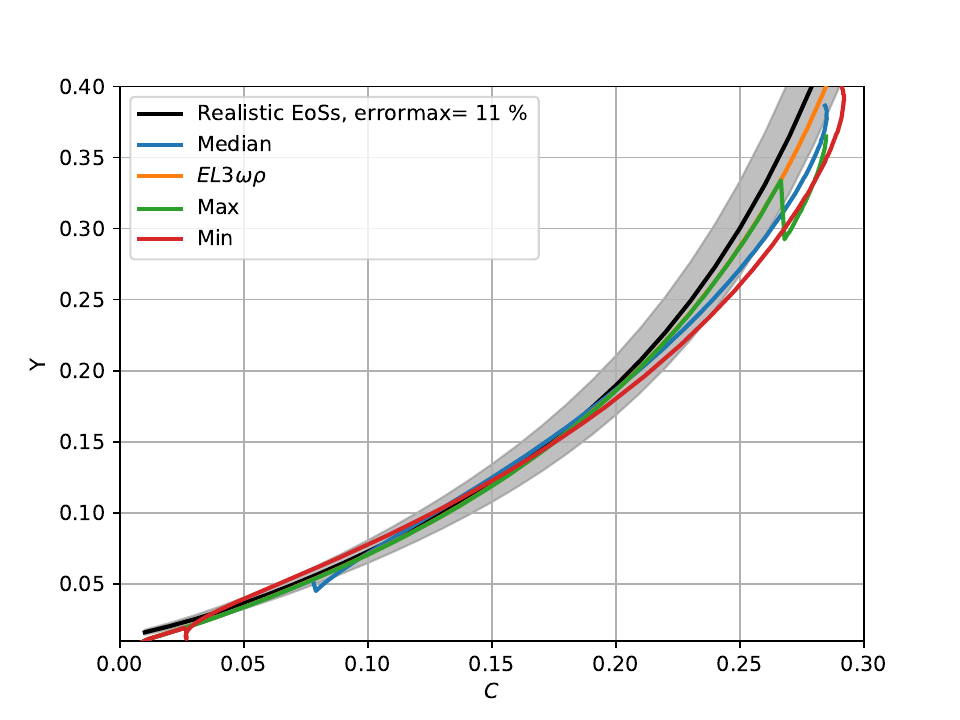}
        \caption{}
        \label{fig4c}
    \end{subfigure}
    \hfill
    \begin{subfigure}[t]{0.49\textwidth}
        \centering
        \includegraphics[scale=0.57]{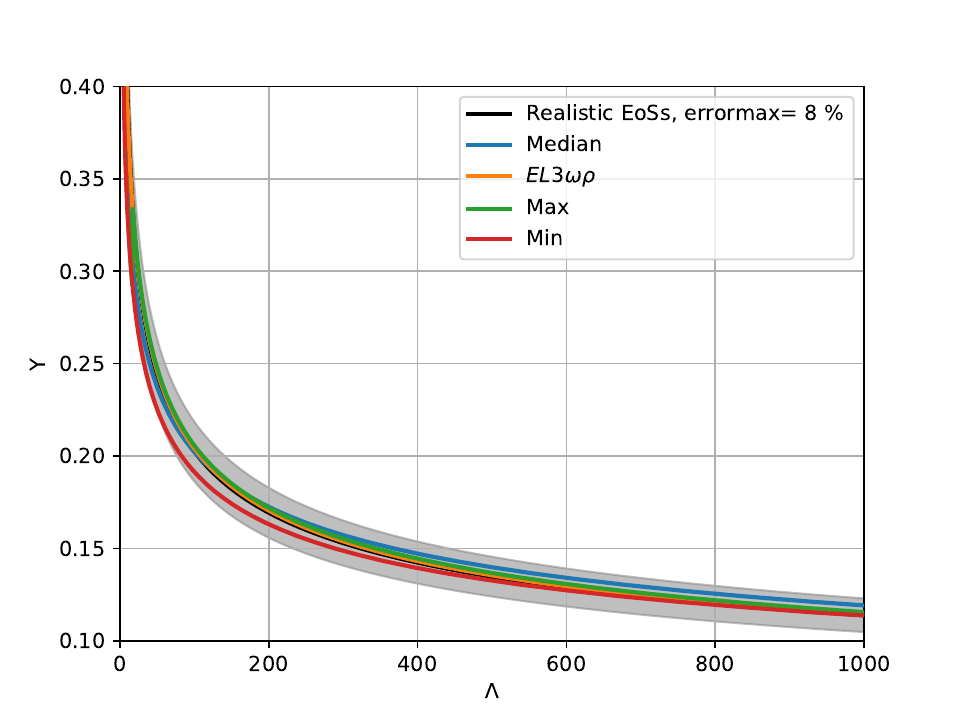}
        \caption{}
        \label{fig4d}
    \end{subfigure}

    \caption{(a) Mass--radius diagram obtained from the vMIT inference for the parameter set $EL3\omega\rho$. The transition density corresponding to the lower bound (min) lies below the imposed numerical cutoff ($> 1.5\,n_0$) and $\mathcal{L}_{\mathrm{phT}}$. The median values (0.27, 154.68, 1.53), represented by the blue solid line, violate the pQCD constraints and yield a transition mass of $\simeq 0.6\,M_{\odot}$. Furthermore, the green solid line (max) results in a small quark core and lies outside the pQCD constraints. The natural log evidence for this model is $-5.46$. (b) Dimensionless tidal deformability $\Lambda$ as a function of the stellar mass. The median value (blue solid line) lies slightly above $\Lambda_{1.4} \approx 400$ and therefore does not satisfy the pQCD-constrained EOS. (c) All inferred results  lie outside the error band for $C > 0.25$. Therefore, estimating the dimensionless moment of inertia from $\Upsilon$ is not recommended. (d) $\Upsilon$ as a function of $\Lambda$ \cite{Saes:2021fzr}. In this case, all inferred results fall within the error band.}
    \label{fig4}
\end{figure}

In contrast to the two-parameter case, the results presented in Fig. 4 are not suitable, since in Fig. \ref{fig4a} the median inferred value violates the pQCD constraints. In Fig.~\ref{fig4b}, all results lie within the observational constraints on $\Lambda_{1.4}$. In Fig.~\ref{fig4c}, the results obtained using the $\Upsilon$--$C$ approximate universal relation lie outside the error band for $C > 0.25$. Furthermore, in Fig \ref{fig4d}, all inferred results fall within the error band.

The results for the two-parameter set $EL3\omega\rho Y$ are presented in Fig \ref{figY1} and Fig. \ref{fig6}.
\begin{figure}[H]
    \centering
    \includegraphics[scale=0.5]{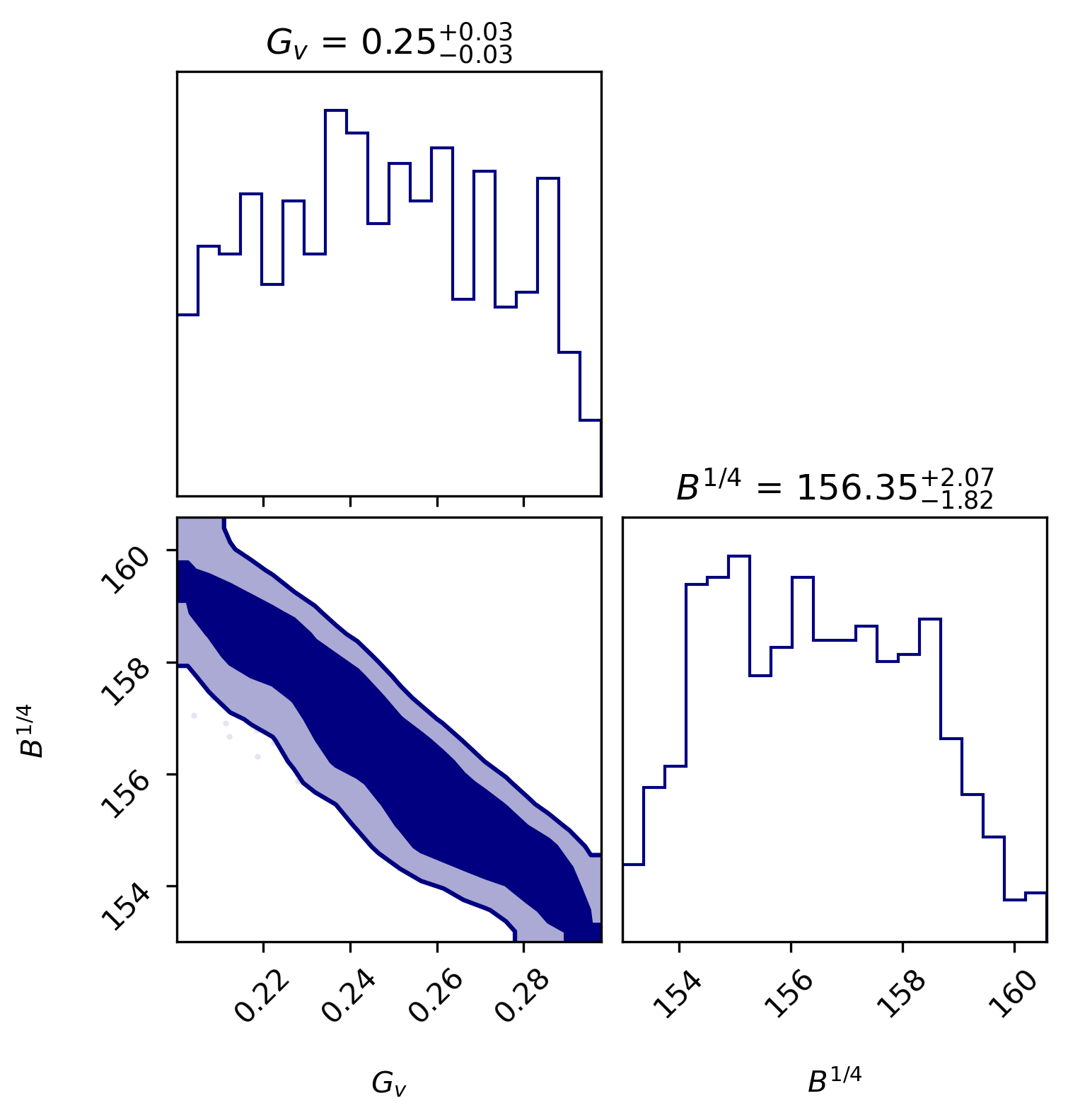}
    \caption{The results for the inference. For quark matter we have used the two vMIT parameters with (0.25, 156.35). For hadronic matter we use the $EL3\omega\rho Y$ parametrization.}
    \label{figY1}
\end{figure}

\begin{figure}[H]
    \centering    
    \begin{subfigure}[t]{0.47\textwidth}
        \centering
        \includegraphics[scale=0.42]{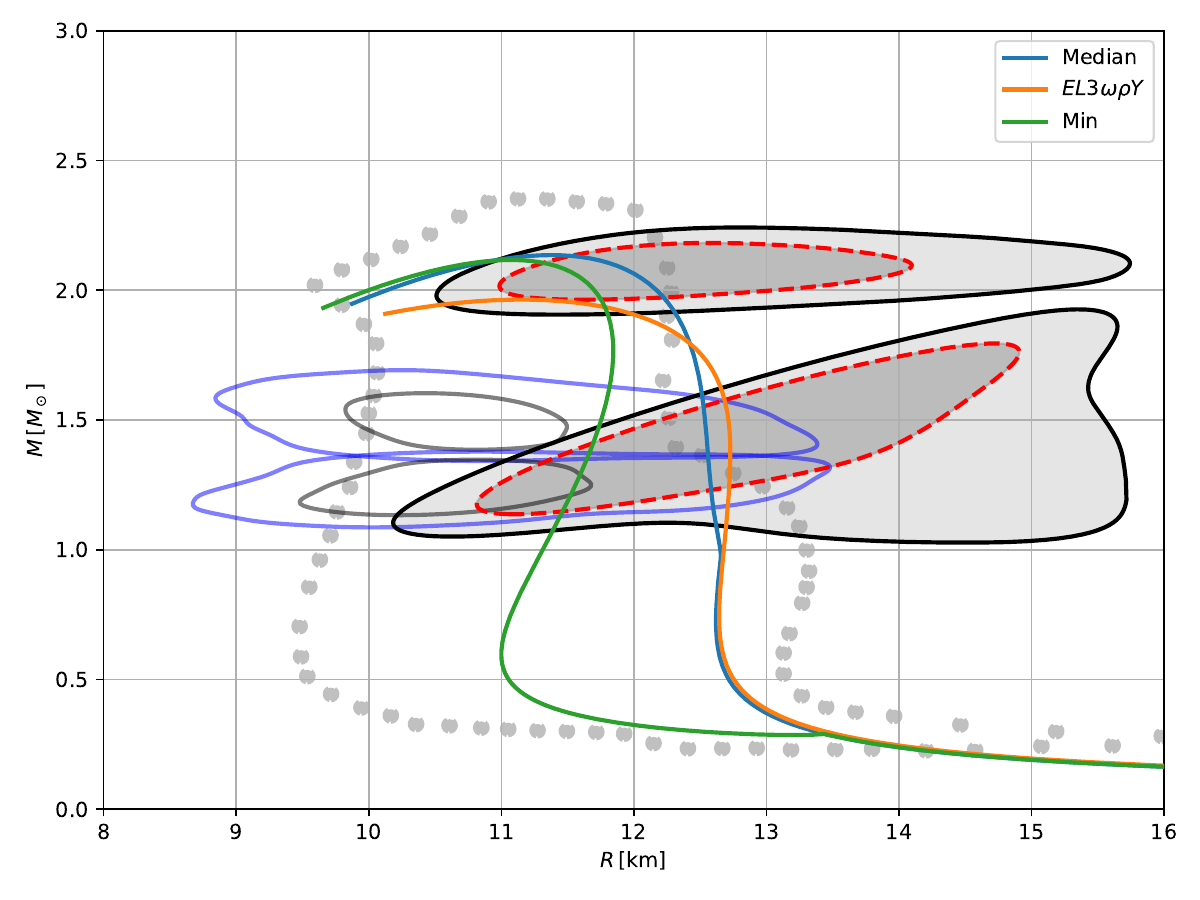}
        \caption{}
        \label{fig6a}
    \end{subfigure}
    \hfill
    \begin{subfigure}[t]{0.47\textwidth}
        \centering
        \includegraphics[scale=0.47]{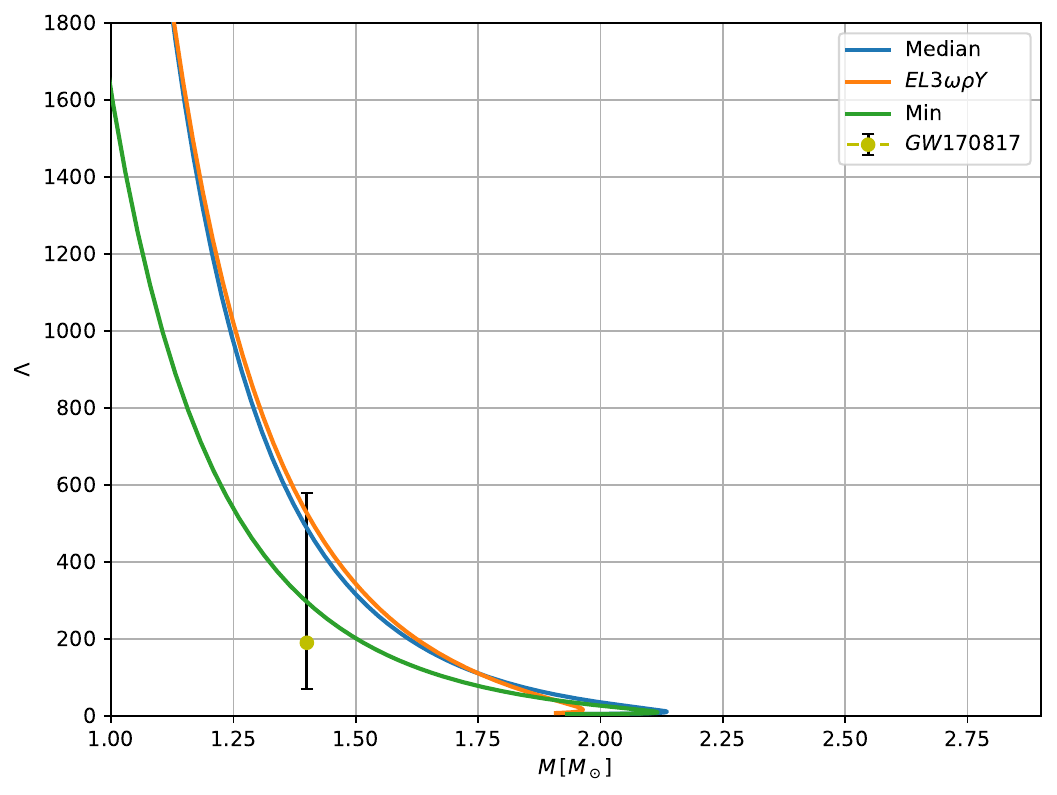}
        \caption{}
        \label{fig6b}
    \end{subfigure}

    \vspace{0.5cm}

    \begin{subfigure}[t]{0.49\textwidth}
        \centering
        \includegraphics[scale=0.57]{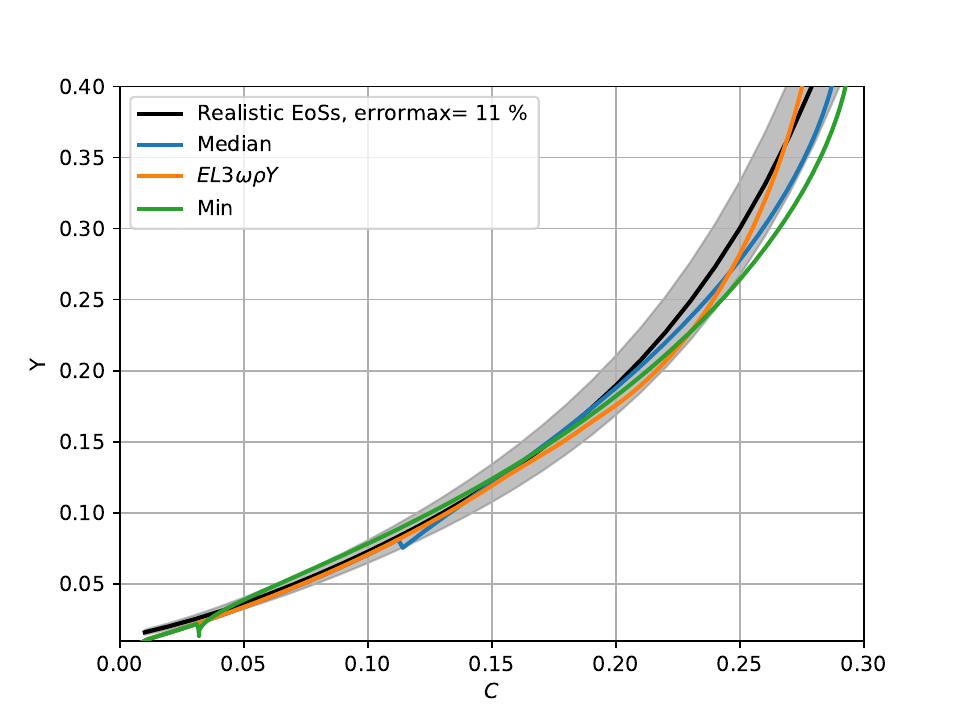}
        \caption{}
        \label{fig6c}
    \end{subfigure}
    \hfill
    \begin{subfigure}[t]{0.49\textwidth}
        \centering
        \includegraphics[scale=0.57]{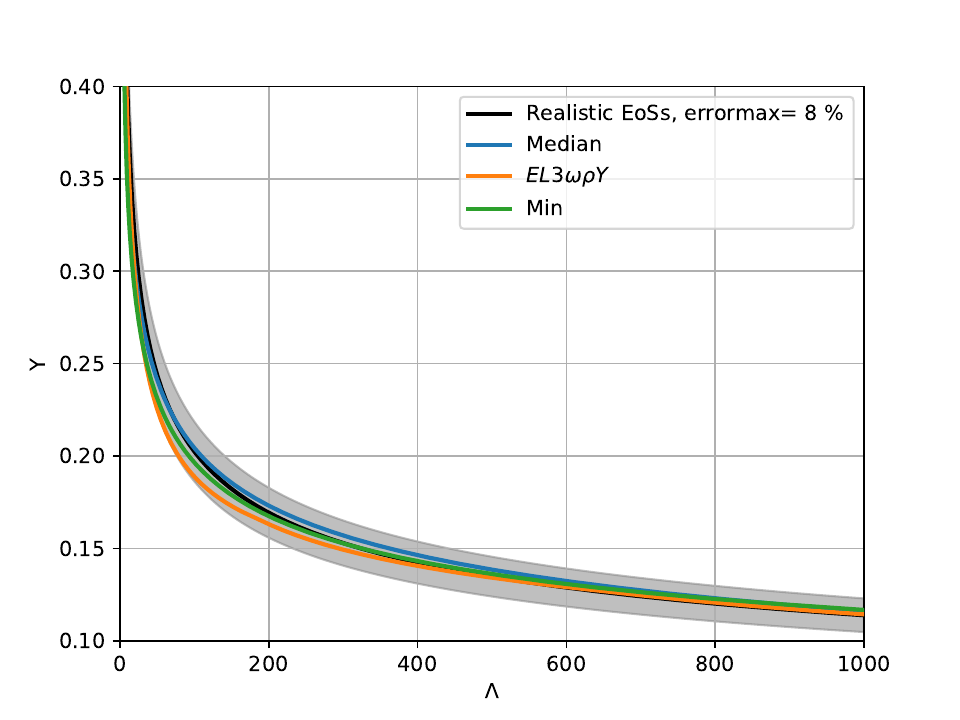}
        \caption{}
        \label{fig6d}
    \end{subfigure}

    \caption{(a) Mass--radius diagram obtained from the vMIT inference for the parameter set $EL3\omega\rho Y$. The transition density corresponding to the lower bound (median) lies below the imposed numerical cutoff ($> 1.5\,n_0$), and no hybrid star is obtained for the upper bound (maximum). The median values $(0.25,\,156.35)$, represented by the blue solid line, violate the pQCD constraints and yield a transition mass of approximately $1\,M_{\odot}$. The natural logarithm of the evidence for this model is $-3.668$. (b) Dimensionless tidal deformability $\Lambda$ as a function of the stellar mass. The median result (blue solid line) yields $\Lambda_{1.4} = 488$ and is therefore inconsistent with the pQCD-constrained EOS. (c) Only the median values (solid blue line) lie within the error band; therefore, the estimate of the dimensionless moment of inertia from $\Upsilon$ is reliable only in this case. (d) $\Upsilon$ as a function of $\Lambda$ \cite{Saes:2021fzr}. In this case, all inferred results fall within the error band. }
    \label{fig6}
\end{figure}

In Fig. \ref{fig6a} the inferred values (0.25, 156.35) for the two-parameter set $EL3\omega\rho Y$ violate the pQCD constraints. The median value of $\Lambda_{1.4}$ in Fig.~\ref{fig6b} exceeds 400 and is therefore outside the pQCD-constrained limit. On the other hand, the approximate universal relations in Figs. \ref{fig6c} and \ref{fig6d} lie within the error band. When comparing the inferred values for the vMIT model with $EL3\omega\rho$ and $EL3\omega\rho Y$, the two-parameter case without hyperons satisfies all constraints, and the universal relations can therefore be used to estimate the dimensionless moment of inertia from $\Upsilon$.

The following results in Fig. \ref{figc1a}, \ref{figc1b}, \ref{fig61} and \ref{fig7} are for the parametrization $NL3^{*}\omega\rho$ with two and three parameters.

\begin{figure}[H]
    \centering
        \centering
        \includegraphics[scale=0.5]{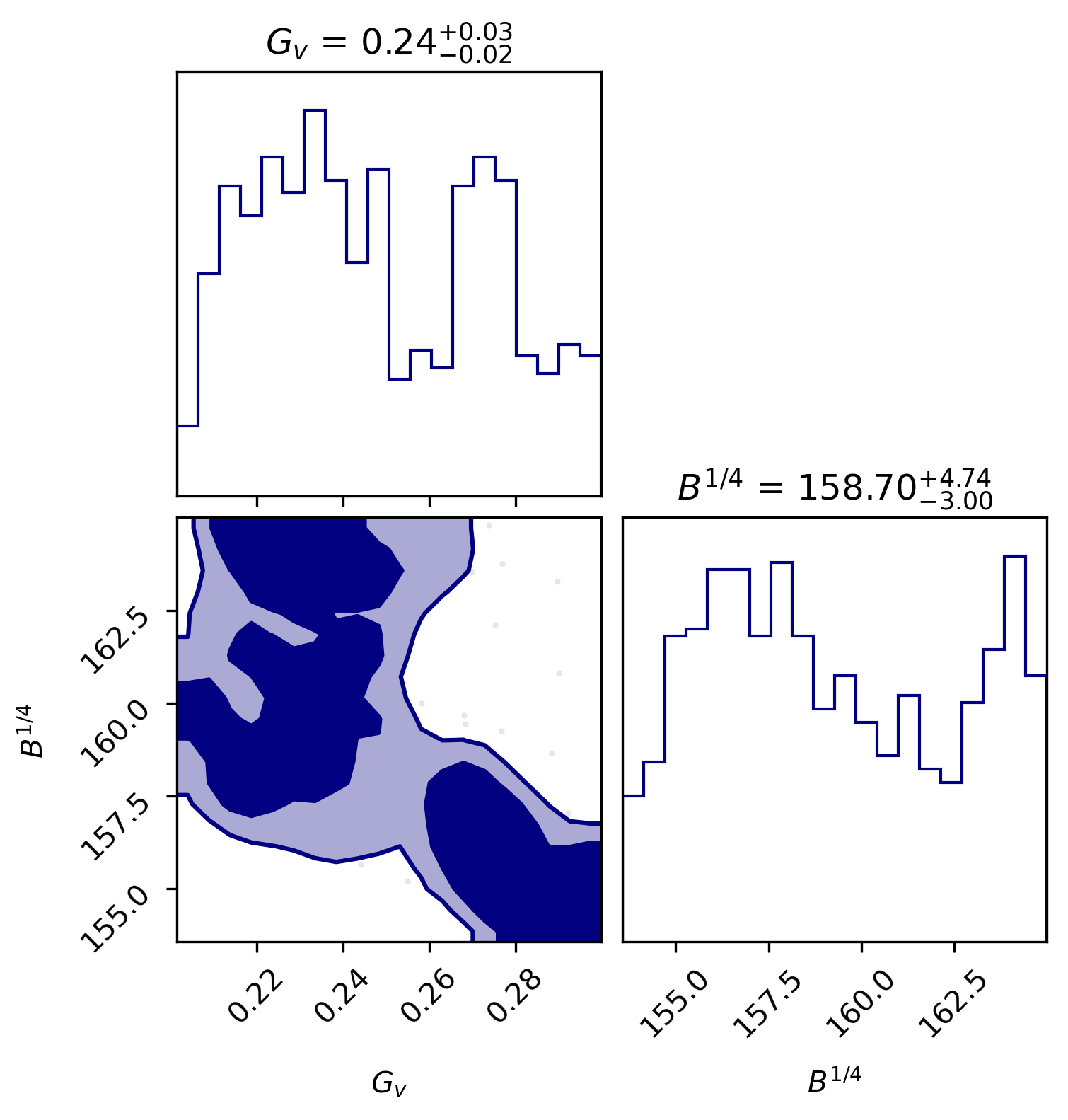}
        \caption{The results for the inference. For quark matter we have used the two vMIT parameters with (0.24, 158.70). For hadronic matter we use the $NL3^{*}\omega\rho$ parametrization.}
        
        \label{figc1a}
 \end{figure}
 
   \begin{figure}[H]
        \hspace{5.7cm}
        \includegraphics[scale=0.5]{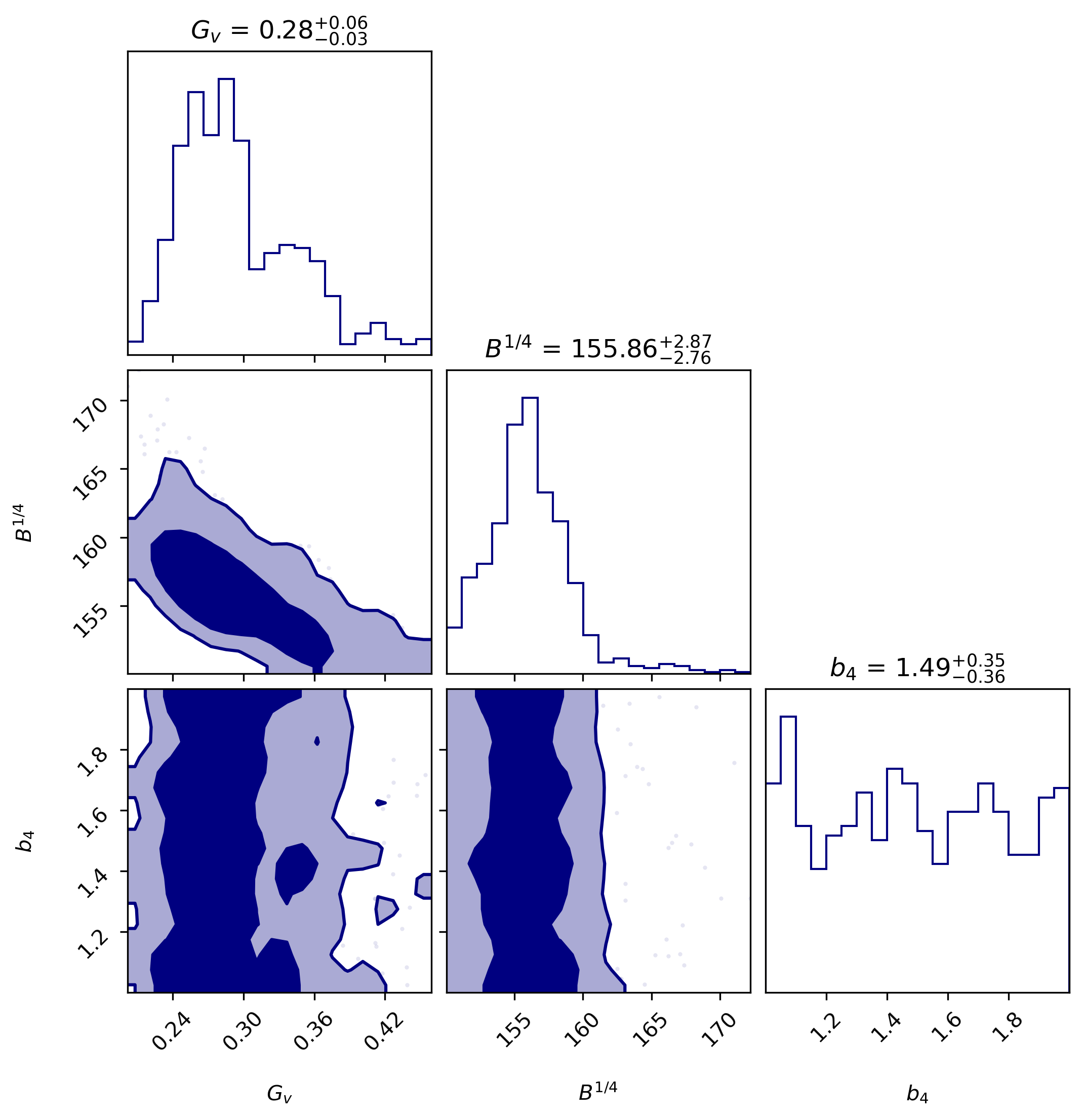}
        \caption{The results for the inference. For quark matter we have used the three vMIT parameters with ((0.28, 155.86,1.49). For hadronic matter we use the $NL3^{*}\omega\rho$ parametrization.}
        \label{figc1b}
    \end{figure}
The dark-to-light blue contours in Figs. \ref{figc1a} and \ref{figc1b} represent the 68$\% $ credible region and  $95\%$ credible region, respectively.

\begin{figure}[H]
    \centering

    \begin{subfigure}[t]{0.47\textwidth}
        \centering
        \includegraphics[scale=0.42]{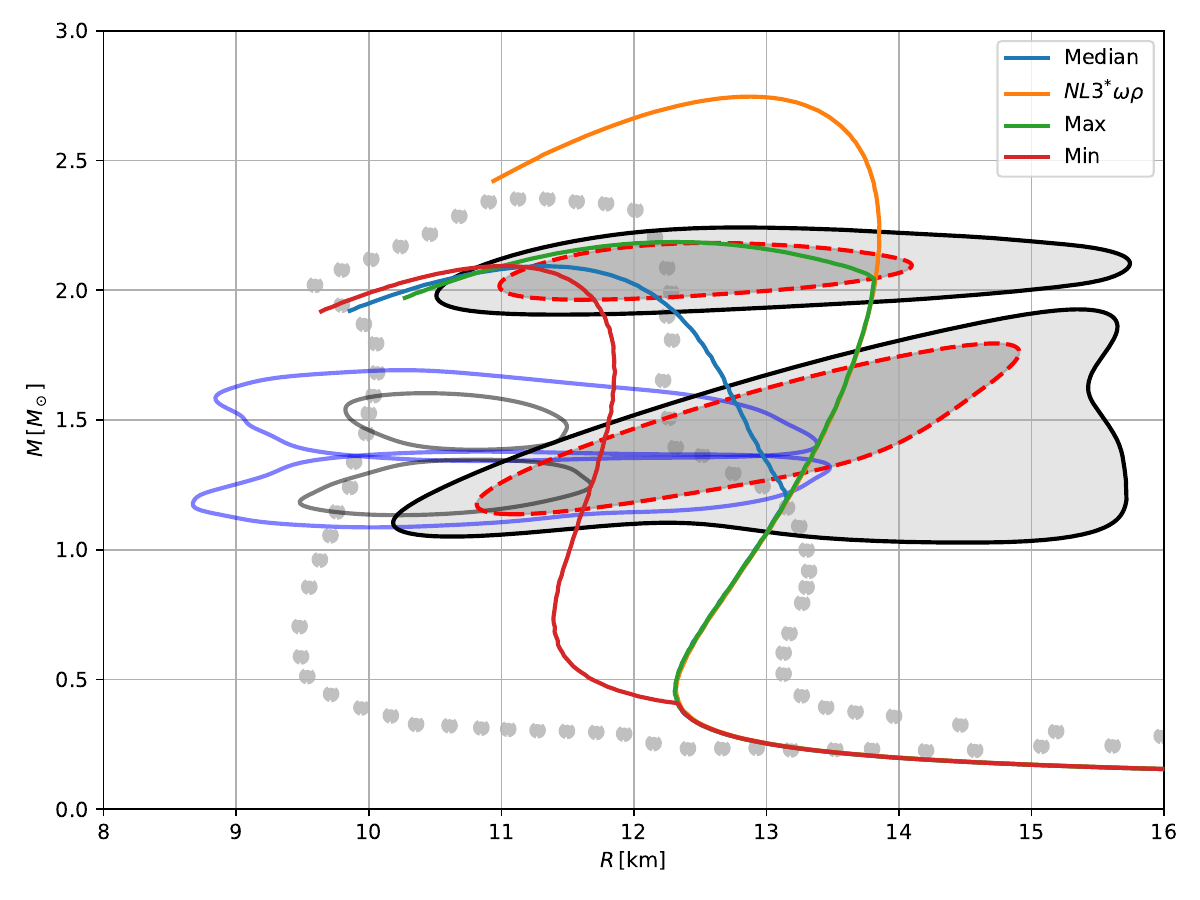}
        \caption{}
        \label{fig61a}
    \end{subfigure}
    \hfill
    \begin{subfigure}[t]{0.47\textwidth}
        \centering
        \includegraphics[scale=0.42]{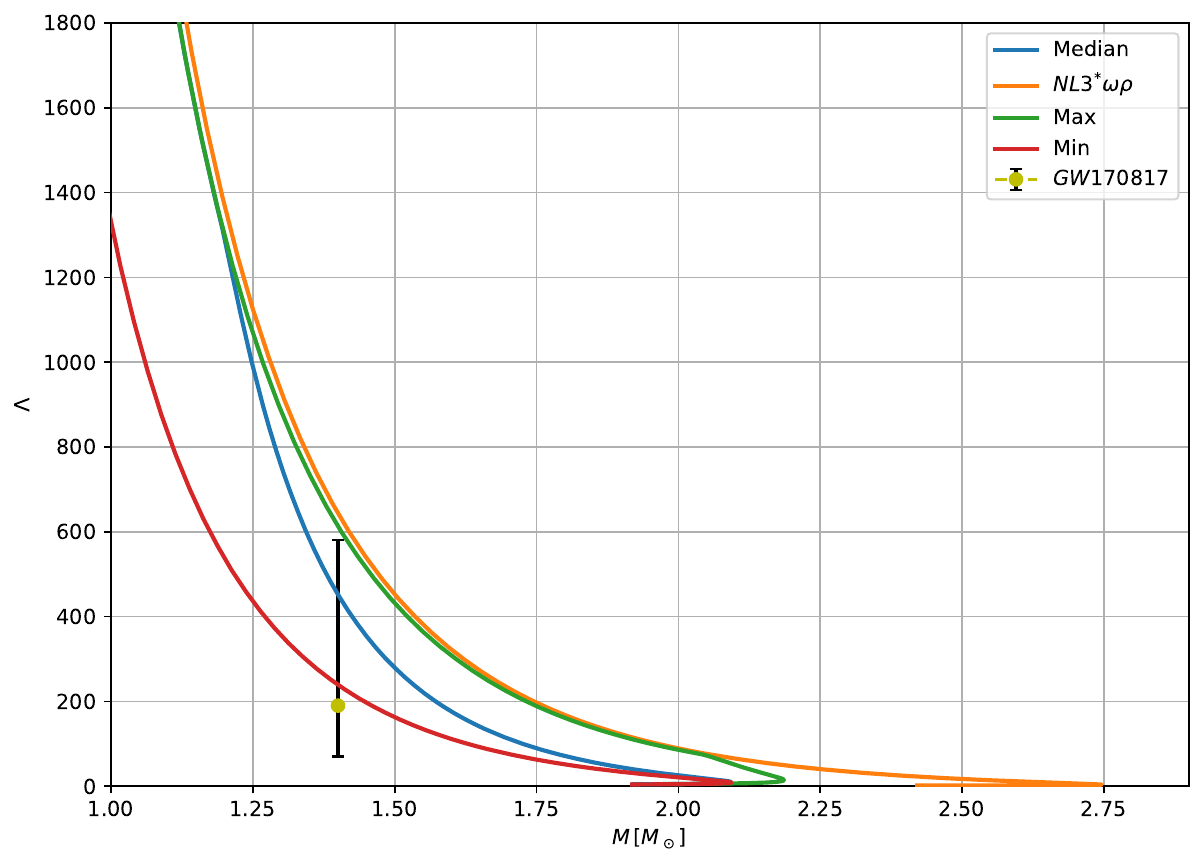}
        \caption{}
        \label{fig61b}
    \end{subfigure}

    \vspace{0.5cm}

    \begin{subfigure}[t]{0.49\textwidth}
        \centering
        \includegraphics[scale=0.57]{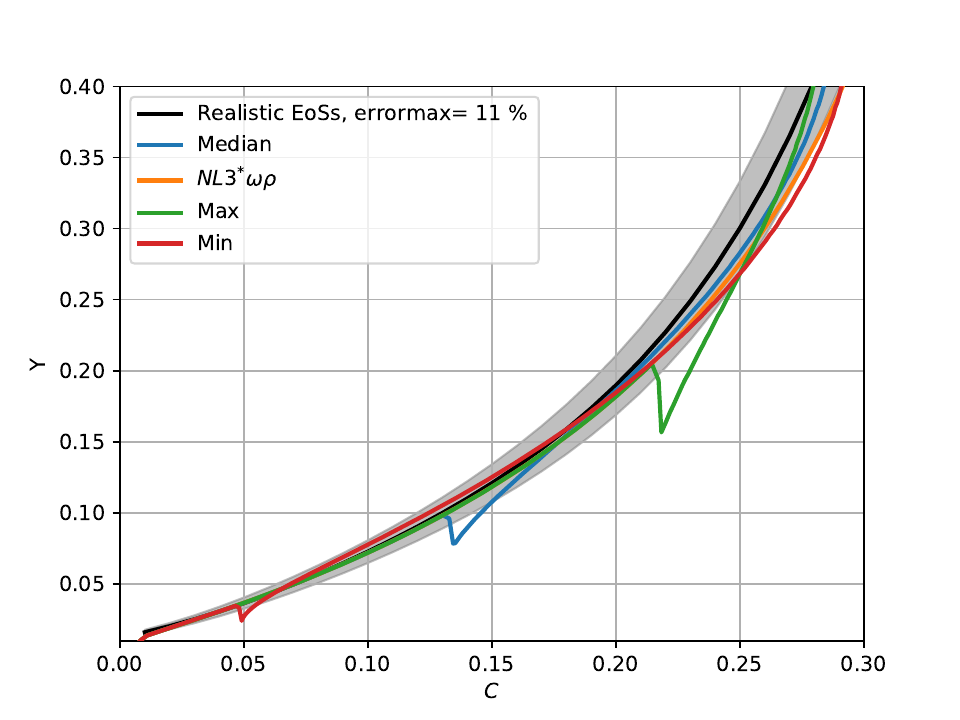}
        \caption{}
        \label{fig61c}
    \end{subfigure}
    \hfill
    \begin{subfigure}[t]{0.49\textwidth}
        \centering
        \includegraphics[scale=0.57]{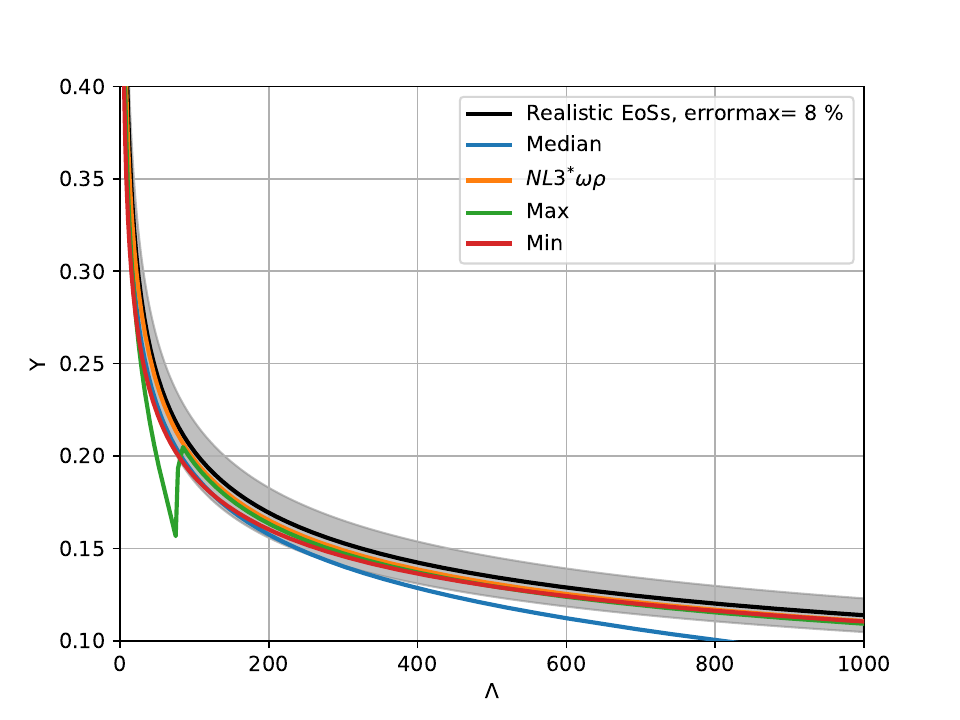}
        \caption{}
        \label{fig61d}
    \end{subfigure}

    \caption{(a) Mass--radius diagram obtained from the vMIT inference for the parameter set $NL3^{*}\omega\rho$. The transition density corresponding to the lower bound (min) lies below the imposed numerical cutoff ($>1.5\,n_0$) and $\mathcal{L}_{\mathrm{phT}}$. Furthermore, the green solid line (max) results in a small quark core and lies outside the pQCD constraints. The median values $(0.24,\,158.70)$, represented by the blue solid line, violate the pQCD constraints and yield a transition mass $\simeq 1.2\,M_{\odot}$. The natural log evidence for this model is $-2.623$. (b) Dimensionless tidal deformability $\Lambda$ as a function of the stellar mass. The median value (blue solid line) lies slightly above $\Lambda_{1.4} > 400$. Thus, it does not lie within the pQCD-constrained EOS. (c) Only the median values (solid blue line) lie within the error band, except at the phase transition; therefore, the estimation of the dimensionless moment of inertia from $\Upsilon$ is only partially reliable in this case. (d) $\Upsilon$ as a function of $\Lambda$ \cite{Saes:2021fzr}. The mean, minimum and maximum values lie outside the error band. }
    \label{fig61}
\end{figure}

In Fig. \ref{fig61a}, the median values violate the pQCD constraint and yield a mass transition of $\simeq 1.2\,M_{\odot}$. In addition, the value of $\Lambda_{1.4}$ corresponding to the median values in Fig.~\ref{fig61b} do not satisfy the pQCD constraint. In Fig.~\ref{fig61c}, only the cusp (solid blue line) lies within the error band. In Fig. \ref{fig61d}, none of the results related to Bayes's inference lies within the error band. Thus, the estimation of the dimensionless moment of inertia from $\Upsilon$ is not reliable, since we require that both approximate relations, $\Upsilon$--$C$ and $\Upsilon$--$\Lambda$, hold simultaneously.

The last results presented in Fig. \ref{fig7} correspond to the same parametrization as before, but with three vMIT parameters.

\begin{figure}[H]
    \centering

    \begin{subfigure}[t]{0.47\textwidth}
        \centering
        \includegraphics[scale=0.42]{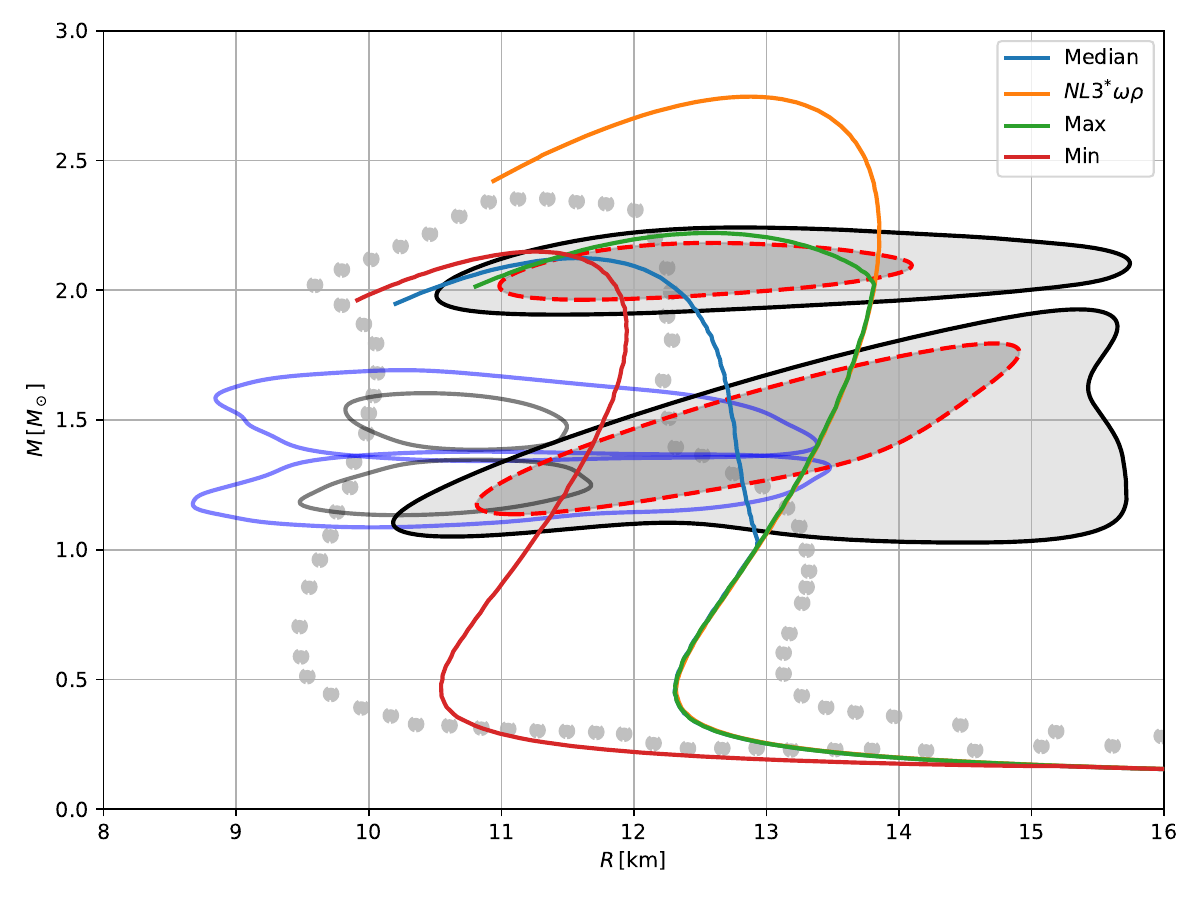}
        \caption{}
        \label{fig7a}
    \end{subfigure}
    \hfill
    \begin{subfigure}[t]{0.47\textwidth}
        \centering
        \includegraphics[scale=0.42]{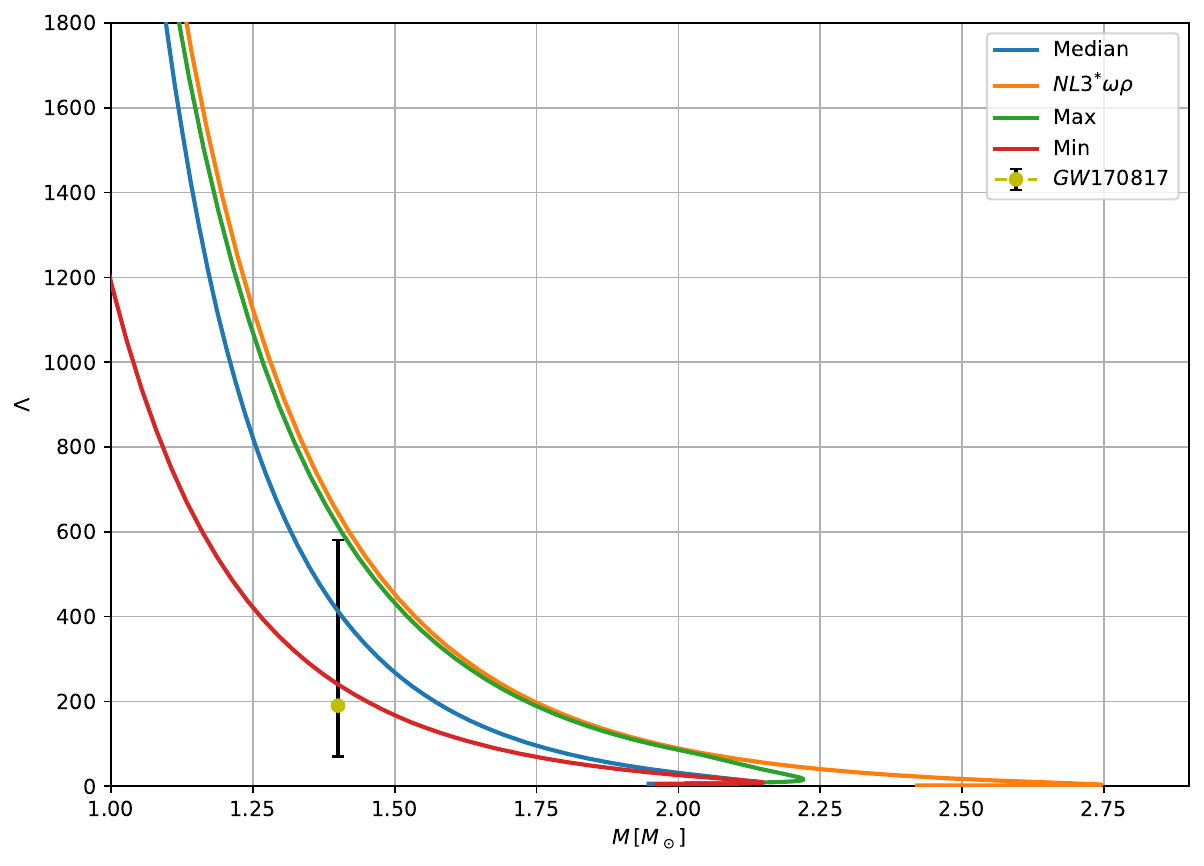}
        \caption{}
        \label{fig7b}
    \end{subfigure}

    \vspace{0.5cm}

    \begin{subfigure}[t]{0.49\textwidth}
        \centering
        \includegraphics[scale=0.57]{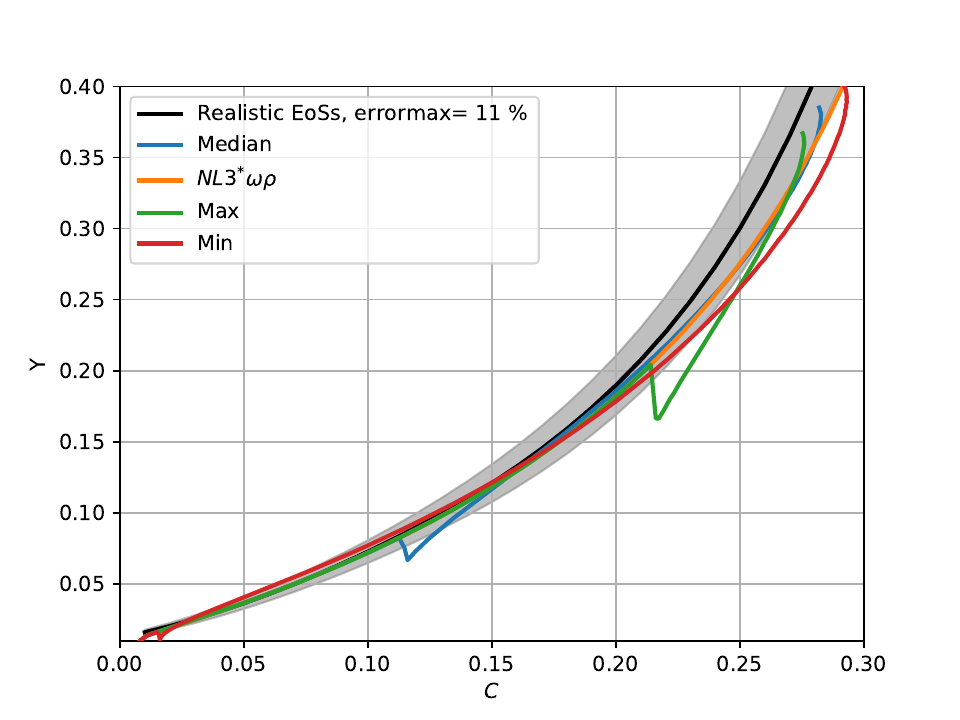}
        \caption{}
        \label{fig7c}
    \end{subfigure}
    \hfill
    \begin{subfigure}[t]{0.49\textwidth}
        \centering
        \includegraphics[scale=0.57]{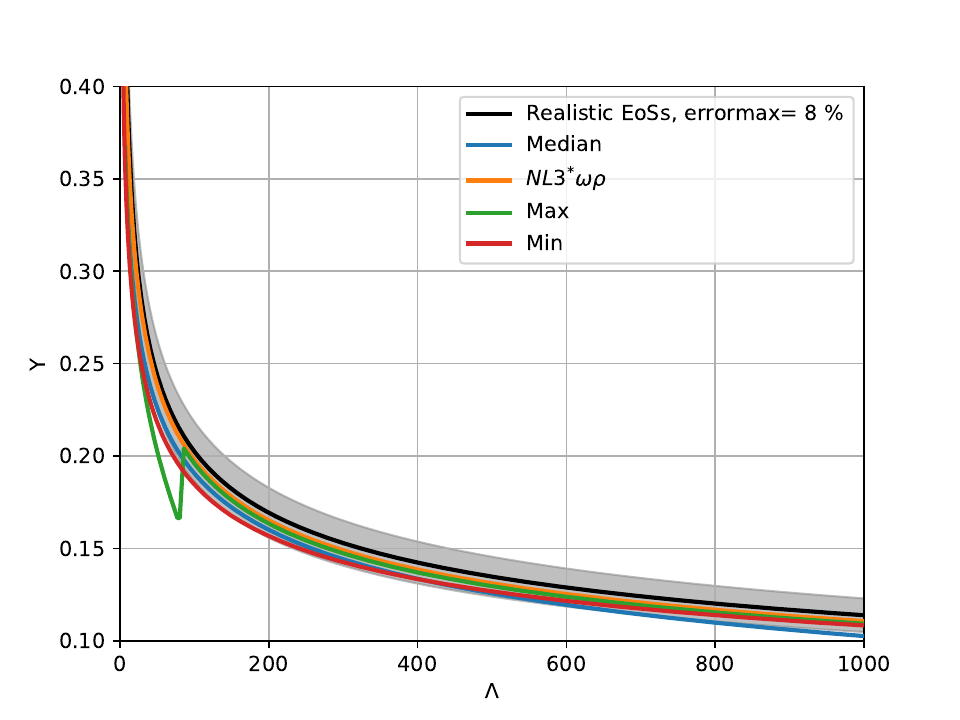}
        \caption{}
        \label{fig7d}
    \end{subfigure}

    \caption{(a) Mass--radius diagram obtained from the vMIT inference for the parameter set $NL3^{*}\omega\rho$. The transition density corresponding to the median values lies below the imposed numerical cutoff (>$1.5\,n_0$) and below $\mathcal{L}_{\mathrm{phT}}$. Furthermore, the green solid line (maximum values) produces only a small quark core and lies outside the pQCD constraints. The median values $(0.28,\,155.86,\,1.49)$, represented by the blue solid line, also violate the pQCD constraints and yield a transition mass of approximately $1.0\,M_{\odot}$. The natural logarithm of the evidence for this model is $-4.170$. (b) Dimensionless tidal deformability $\Lambda$ as a function of stellar mass. The median result (blue solid line) lies slightly above $\Lambda_{1.4} \approx 400$ and is therefore inconsistent with the pQCD-constrained EOS. (c) Only the median values (solid blue line) lie within the error band, except near the phase transition; therefore, the estimate of the dimensionless moment of inertia from $\Upsilon$ is only partially reliable in this case. (d) $\Upsilon$ as a function of $\Lambda$ \cite{Saes:2021fzr}. The mean, minimum and maximum values lie outside the error band.   }
    \label{fig7}
\end{figure}

In Fig. \ref{fig7a}, the median values violate the pQCD constraint and yield a mass transition of $\simeq 1.0\,M_{\odot}$. In addition, the value of $\Lambda_{1.4}$ corresponding to the median values in Fig.~\ref{fig7b} does not satisfy the pQCD constraint. In Fig.~\ref{fig7c}, only the cusp (solid blue line) lies within the error band. In Fig. \ref{fig7d}, none of the results related to Bayes's inference lies within the error band. Thus, as in the two-parameter case, the estimation of the dimensionless moment of inertia from $\Upsilon$ is not reliable, since both approximate relations, $\Upsilon$--$C$ and $\Upsilon$--$\Lambda$, must hold simultaneously.

Another constraint of interest is that imposed by the pQCD EOS on the $\Lambda_{1} - \Lambda_{2}$. For the $NL3^{*}\omega\rho$ parametrization, with three parameters, we have:

\begin{figure}[H]
    \centering
    \includegraphics[scale=0.6]{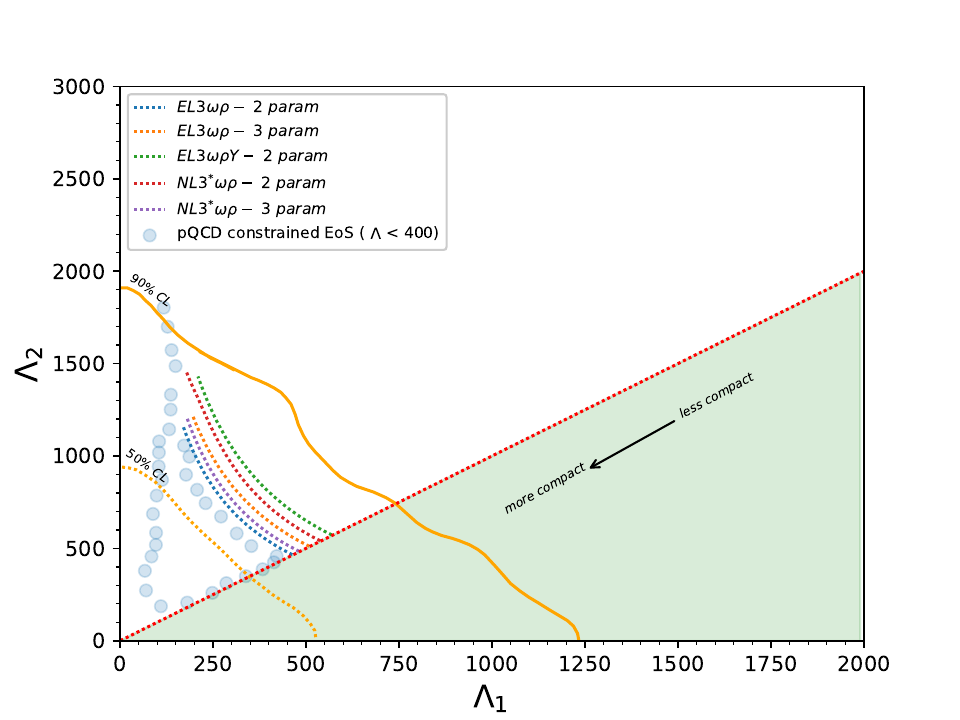}
    \caption{One can note that only the parametrization set $EL3\omega\rho$ with two parameters (dashed blue line) is close to the pQCD constraint and lies within the $90\%$ credible level. }   
    \label{fig:8b}
\end{figure}

In Fig.~\ref{fig:8b}, the dashed blue line is the favored case, as it most closely approaches the pQCD constraint (blue solid dots).

The results for the approximate universal relation $C$--$\Lambda$ \cite{YAGI20171} are shown in Fig.~\ref{fig:123}. It can be seen that all inferred values lie within the stated accuracy of this approximate universal relation.

\begin{figure}[H]
    \centering
    \includegraphics[scale=0.5]{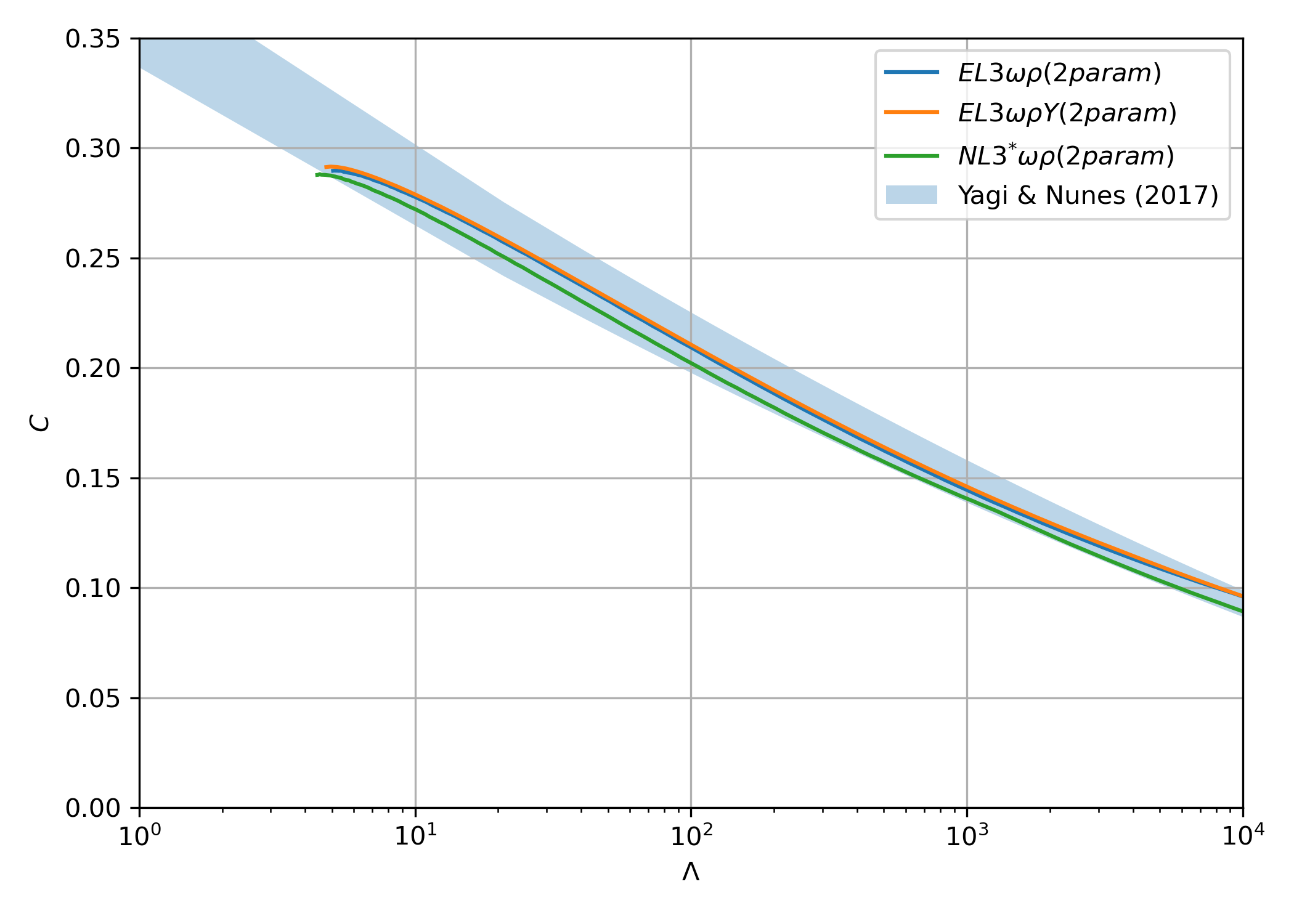}
    \caption{Comparison between the fitted relation and our results for  $EL3\omega\rho$ (0.22, 157.47), $EL3\omega\rho Y$(0.25, 156.35) and   $NL3^{*}\omega\rho$(0.24, 158.70).}   
    \label{fig:123}
\end{figure}

\subsection{Bayes Factor}

The Bayes factor is a method for comparing models based on their evidence. It also reflects Occam’s razor, since models with more parameters are naturally penalized. The Bayes factor is defined as
\begin{equation}
    B_{12} = \frac{\text{evidence of model 1}}{\text{evidence of model 2}} ,
\end{equation}
and the criterion for interpreting $B_{12}$ is given by the Jeffreys scale \cite{Trotta01032008}. For more details see Table \ref{table4}.

\begin{table}[h]
\centering
\begin{tabular}{ccc}
\hline
$\ln B_{12}$ & Probability & Strength of evidence for $M_1$ \\ 
\hline
$<1.0$  & $<0.750$ & Inconclusive \\
$1.0$   & $0.750$  & Weak evidence \\
$2.5$   & $0.923$  & Moderate evidence \\
$>5.0$  & $0.993$  & Strong evidence \\
\hline
\end{tabular}
\caption{Interpretation of Bayes factors according to the Jeffreys scale \cite{Trotta01032008}.}
\label{table4}
\end{table}

For the EOS $EL3\omega\rho$, we compare the case with two parameters (model 1) to the case with three parameters (model 2).
\[
\ln(B_{12}) = -4.54 + 5.46 = 0.92
\]
This result corresponds to \textit{inconclusive} evidence in favor of model 1 ($M_1$).  Furthermore, based on the constraints from pQCD (gray dots) and universal relations, only model 1 is consistent with that constraint. The inconclusive evidence in favor of $M_1$ can be observed when comparing the two sequences of stellar curves (blue solid line). The main difference occurs in the transition mass. Furthermore, the results for the EOS $NL3\omega\rho$ are inconsistent with the pQCD constraints; however, its Bayes factor is 1.547, which indicates weak evidence in favor of the two-parameter model.

The authors of \cite{deMouraSoares:2023cqq} demonstrated that both Gaussian and uniform priors for the different symmetry energy slope parameter interval $L$ yield essentially the same Bayes factor when using the same full likelihood function for neutron stars. According to the Jeffreys scale presented in their Table~3, the evidence against $M_{A}$ is "substantial". The latter conclusion is similar to our results.

\subsection{Phase transition properties and macroscopic features}
In Table \ref{T1}, we present the following parameters related to the beginning of the phase transition: baryon density number ($n_{b}$), critical pressure ($P_{\mathrm{critical}}$), the width of the phase transition in terms of energy density ($\Delta \varepsilon$), and the critical baryon chemical potential ($\mu_{\mathrm{critical}}$).

\begin{table*}[htbp]
    \centering
    \caption{Phase transition properties obtained from Bayesian inference.}
    \label{T1}
    \begin{tabular}{c|c|c|c|c}
    \hline
        \hline
        \textbf{Set} & $n_{b}~[1/\mathrm{fm}^{3}]~(n_{b}/n_{0}),~n_{0}=0.15~[1/\mathrm{fm}^{3}]$ & 
        $P_{\mathrm{critical}}~[\mathrm{MeV}/\mathrm{fm}^{3}]$ & 
        $\Delta \varepsilon~[\mathrm{MeV}/\mathrm{fm}^{3}]$ & 
        $\mu_{\mathrm{critical}}~[\mathrm{MeV}]$ \\ 
        \hline
        (0.22, 157.47) - $EL3\omega\rho$ &  0.28~(1.90) & 18.15 & 61.16 & 1039.50 \\ 
        \hline
        (0.27, 154.68, 1.53) - $EL3\omega\rho$ & 0.26~(1.74) & 13.61 & 51.71 & 1023.41 \\ 
        \hline
        (0.25, 156.35) - $EL3\omega\rho Y$ & 0.32~(2.13)  \footnote{The transition to the deconfined quark phase occurs before the appearance of the $\Lambda^{0}$ hyperon, which takes place at $n_b \approx 0.38 ~\mathrm{fm}^{-3}$. } & 26.04 & 32.16 & 1065.82 \\ 
        \hline
         (0.24, 158.70)  - $NL3^{*}\omega\rho$ & 0.28~(1.87) & 28.21 & 96.50 & 1081.65 \\ 
        \hline
       (0.28, 155.86, 1.49) - $NL3^{*}\omega\rho$ & 0.26~(1.74) &21.90 & 77.15 & 1058.40 \\ 
        \hline
    \end{tabular}    
\end{table*}

The values obtained in Table~\ref{T1} for $n_b^{\text{trans}}$ lie within the likelihood function for the phase transition. The parameter set $(0.25, 156.35)$ --- $EL3\omega\rho Y$ --- is the only one that is not close to the mean value assumed in the likelihood function.

Furthermore, if we assume that the phase transition must occur for $\mu_{B} > 1050~\mathrm{MeV}$, according to the Polyakov loop formalism, only the following inferred values satisfy this condition: $(0.27, 154.68, 1.53)$ --- $EL3\omega\rho$, $(0.24, 158.70)$ --- $NL3^{*}\omega\rho$, and $(0.28, 155.86, 1.49)$ --- $NL3^{*}\omega\rho$. However, these results lie outside the pQCD-constrained EOS.

\begin{table*}[htbp]
    \centering
        \caption{Maximum values for mass, radius, speed of sound, $\Lambda_{1.4}$, and $R_{1.4}$ for the inferred sets.}
    \begin{ruledtabular}

\begin{tabular}{c |c| c| c|c|c}
  Parameter set &$M_{max} [M_\odot] $ & $R_{max} [ km ]$ & $c_{s}^{2}(max)$ & $\Lambda_{1.4} $& $R_{1.4} [ km ]$  \\    
 \hline
(0.22, 157.47) - $EL3\omega\rho$   & 2.07 & 11.10 &  0.48 & 397 & 12.24 \\  
 \hline
  (0.27, 154.68, 1.53) - $EL3\omega\rho$  & 2.12 & 11.46 & 0.43 & 439 & 12.40 \\  
 \hline
 (0.25, 156.35) - $EL3\omega\rho Y$   & 2.13 & 11.36 &  0.50 & 488 & 12.55 \\  
 \hline
 (0.24, 158.70) - $NL3^{*}\omega\rho$   & 2.09 & 11.32 &  0.49 & 452 & 12.93 \\  
 \hline
 (0.28, 155.86, 1.49) - $NL3^{*}\omega\rho$  & 2.12 & 11.55 & 0.43 & 412 & 12.77 \\  
 \end{tabular}
  \end{ruledtabular}
\label{T2}
\end{table*}
Table~\ref{T2} shows similar values, with the exception of $\Lambda_{1.4}$. Additionally, the parameter set obtained through Bayesian inference satisfies the instability condition for matter composed of u, d, and s quarks, as illustrated in Fig. 5(a) of Ref. \cite{Laskos-Patkos:2023tlr}. A similar conclusion is presented in Fig. 4 of Ref.\cite{Lopes:2020btp}. Moreover, when considering the Dirac sea contribution of the quarks in the vector channel, Fig.~7 of Ref. \cite{Lopes:2020btp} demonstrates that the obtained parameters $(G_{v}, B^{1/4}, b_{4})$ are also consistent with the \textit{instability condition}. The inclusion of the Dirac sea contribution in the vector channel, as pointed out by the authors\cite{Lopes:2020btp}, softens the EOS at high densities. Thus, it is possible to have higher values of \( G_{v} \) while at the same time not increasing the baryon chemical potential for a fixed bag value.

The density transition found for the $EL3\omega\rho Y$ is before the appearance of $\Lambda^{0}$, as we can see in Fig. \ref{fig_mb} in the Appendix.

The authors \cite{Raithel_2018}  found that the tidal deformability parameter, $\Lambda$, is approximately independent of the component masses when the chirp mass is fixed in neutron star mergers. Moreover, this result was obtained in the quasi-Newtonian limit, owing to its simplicity and its similarity to the full general relativistic treatment. For the chirp mass measured from GW170817, this leads to at most a $4\%$ correction to the effective tidal deformability across the full range of mass ratios. Thus, their results imply that, for the event GW170817, the radius of a $1.4 ~M_\odot$ neutron star satisfies $R_{1.4} \sim 13 ~\text{km}$ at the $90\%$ confidence level. Our results for $R_{1.4}$, presented in Table~\ref{T2}, agree with their findings.

\subsection{Stability Against Radial Oscillations of Hybrid Stars}

It is well known that the existence of a phase transition inside the neutron star could alter its stability properties. In fact there are two conditions to study the stability of a star. The first is the static stability condition, related to $\partial M/\partial  \epsilon_c\geq 0$, and the second case is the dynamical stability condition related to $\omega > 0$, where $\omega$ is the frequency of the oscillation \cite{2018ApJ...860...12P}.

In this work we have considered the radial oscillations of hybrid stars, whose core is composed of deconfined quark matter, and the external envelope is formed by hadronic matter. This two layer structure is produced by a phase transition inside the star. In principle we can classify those transitions as rapid or slow, with each type having a different effect on the oscillatory properties of the star. In the first case, i.e. rapid transitions, the usual static stability condition $\partial M/\partial \epsilon_c\geq 0$, where $\epsilon_c$ is the central density of a star whose total mass is $M$, remains always true. But in the case of a slow transition, the frequency that belongs to the fundamental mode can be a finite real number ($\omega > 0$), which means stability, even for branches of the stellar models that verify the criterium $\partial M/\partial \epsilon_c \leq 0$ (this generally happens after the point of maximum mass on the diagram $M$ vs $R$) \cite{2018ApJ...860...12P,2012IJMPS..18..105V, Mariani:2024gqi}. 

For all of our models, considering slow phase transitions, we verified that beyond the point of maximum mass in the mass-radius diagrams, there exists a very small stellar branch satisfying $\omega > 0$. Therefore, in practical terms, the two conditions agree very well.

\section{Conclusions} \label{conclusions}

We have investigated the impact of the phase transition from nuclear matter to deconfined quark matter inside hybrid stars. For this objective we performed a Bayesian inference using the observational data in Table~\ref{tab1}. This step is important in order to come up with the likelihood functions. The uniform prior intervals were defined based on values reported in the literature\cite{Lopes:2020btp}. To model a first-order transition, the Maxwell construction was applied. The outer crust is modeled with the BPS EOS\cite{bps}.  \newline  

It is important to note that only the combination $BPS$--$EL3\omega\rho$, applying the vMIT model with two parameters, lies within the uncertainties calculated with $\chi$EFT for densities below $1.1n_{0}$, satisfies the universal relations $\Upsilon$--$C$ and $\Upsilon$--$\Lambda$, and is close to the critical chemical potential predicted by the Polyakov loop formalism. Furthermore, the transition mass is found to be $\simeq 0.8,M_{\odot}$, and the transition density lies within the range expected for hybrid stars \cite{Albino:2024ymc}. Our maximum mass and $R_{1.4}$ lie within the ranges reported in Table~I of \cite{PhysRevLett.120.172703} for four tropes ($\Lambda_{1.4} < 400$).

In addition, we have assumed that results which violate the pQCD-constrained EOS are not suitable, according to our previous premises. 

An additional test was implemented for all hybrid stars, i.e., radial oscillations for both rapid and slow transitions, and it was found that these stars satisfied both conditions. \newline 

In Ref. \cite{jia}, the authors studied twin- and triplet-star configurations for pure nuclear matter, nuclear matter with hyperons, and $\Delta$-resonance--hyperon admixed matter. Their hadronic equation of state  was calculated within the RMF framework using the DD-ME2 parametrization, which employs density-dependent meson--nucleon couplings. The hyperonic EOS was constructed as an extension of DD-ME2. For quark matter, the EOS was modeled using a synthetic constant-sound-speed (CSS) parametrization.

In that work, the maximum mass of the hybrid star was fixed at two solar masses, while the transition mass was varied. In Table II of Ref. \cite{jia}, for the case of a single phase transition leading to twin-star configurations, it is possible to compare our results for the transition mass and transition density in both scenarios: with a purely nucleonic envelope and with a nucleonic-plus-hyperonic envelope. Our result for nuclear matter with $(0.22, 157.47) - EL3\omega\rho$ is comparable to their result for mass transition = 0.8 $~M_{\odot}$ and $nb/n0$ =1.734. However, they found for nuclear-plus-hyperonic envelope, mass transition = 1.6 $~M_{\odot}$ and $nb/n0$ = 2.549. Both values are greater than $EL3\omega\rho Y$. The maximum mass for $EL3\omega\rho Y$ is $2.13~M_{\odot}$ while for the work of Ref. \cite{jia} it is $2.0~M_{\odot}$. 

Our result for the vMIT set $(0.22, 157.47)$ obtained with the $EL3\omega\rho$ parametrization is very similar to that reported in Ref. \cite{Albino:2024ymc} in terms of $M_{\mathrm{max}}$, $nb_{\mathrm{trans}}$, $R_{\mathrm{max}}$, $c_s^2(\mathrm{max})$, and $R_{1.4}$. For the hadronic envelope, the authors also employed the relativistic mean-field  framework using the BMPF220  parametrization (soft EOS), while the quark matter phase was described within the mean-field theory of QCD.

In Ref. \cite{kumar}, the authors also employed the vMIT model for the quark phase, considering a bag that depends on the baryon density. Their results in Table II for the density-dependent bag case with parameters $(0.35, 172)$ are close to our results obtained with the $EL3\omega\rho$ parametrization and parameters $(0.22, 157.47)$ for the following quantities: $\mu_t$, $P_c$, $R_{2.08}$, $\Lambda_{1.4}$, $mass_{trans}$ and $\Delta R_{1.4} < 0.5\mathrm{km}$. The hadronic envelope was modeled using the Next-to-next-to-leading order and density-dependent meson-exchange (DD-MEX) parametrization within the relativistic mean-field framework.

In Ref. \cite{Podder:2023dey}, hybrid stars were constructed using the MIT bag model with a density-dependent bag constant and the RMF parametrization NL3$\omega\rho$6. The latter equation of state predicts a neutron star maximum mass of $M_{\mathrm{max}} \sim 2.75,M_{\odot}$. Their hybrid star models also exhibit a mass transition similar to the one found in our results. However, their model does not provide a satisfactory description of the observational constraints from PSR J0030+0451. A soft or intermediate hadronic EOS could produce results similar to ours in the case of the $EL3\omega\rho$ parametrization.

The authors of \cite{sophia_quark_c} also analyzed the phase transition in hybrid stars using Bayesian inference. For the hadronic envelope, they used the quark mean field (QMF) model \cite{Zhu:2018ona}, which provides a soft EOS, while the quark matter phase was described using the CSS parametrization. Their results also presented a large quark core. For their hybrid stars, considering the \(90\%\) confidence interval from the combined GW170817 + NICER analysis, they obtained the following results:
$nb_{trans}/n_{0}$ $2.04^{+1.22}_{-0.83}$, $c_{s}²(max)=0.81^{+0.17}_{-0.28}$, $M_{max}=2.36^{+0.49}_{-0.26} M_{\odot}$, $R_{max}=10.96^{+1.79}_{-1.01} km$, $R_{1.4}=11.70^{+0.85}_{-0.74} km$, $R_{2.0}=11.66^{+1.46}_{-1.22} km$ and   $\Lambda_{1.4}=312^{+254}_{-124}$. These results are in good agreement with those obtained in our best-case inference, \((0.22, 157.47)\) -- EL3\(\omega\rho\).

Our results favor a hadronic envelope described by a soft EOS, with the quark matter phase modeled using the vMIT bag model. However, the vMIT bag model presents intrinsic limitations, including the absence of the non-conformal behavior expected in pQCD and the lack of a kink in the EOS of the envelope, which, according to the authors of Ref.~\cite{Annala:2019puf}, is necessary for the formation of hybrid stars (quark cores). In addition, as mentioned before, our results are very similar to those reported in the literature using both Bayesian inference and heuristic approaches.

\section*{Acknowledgments}
This work is a part of the project INCT-FNA proc. No. 408419/2024-5, and also supported by Conselho Nacional de Desenvolvimento Cient\'ifico e Tecnol\'ogico (CNPq) under Grant No. 303490/2021-7 (D. P. M.). F. K. would like to thank FAPESC/CNPq for financial support under Grants No. 174332/2023-8. C. V. F. acknowledges the financial support of the productivity program
of the CNPq, with Project
No. 304569/2022-4. C. H. L. is thankful to the São Paulo Research Foundation FAPESP (Grant
No. 2020/05238-9), to CNPq (Grants No. 401565/2023-8, 409736/2025-2 and No. 305327/2023-2).

\clearpage
\appendix
\section{Hyperon-Meson Coupling}

The hyperons included in the \( EL3\omega\rho Y \) model are \( \Lambda, \Sigma^-, \Sigma^0, \Sigma^+, \Xi^-, \text{and}~ \Xi^0 \). Only one hyperon-meson coupling is well established, which is related to \( \Lambda^0 \) and has a potential depth of \( U_\Lambda = -28~{\rm MeV} \). For the remaining hyperons, the vector mesons are constrained by the SU(6) symmetry group \cite{RevModPhys.38.215}. However, for the scalar mesons \cite{Lopes:2020rqn,Lopes:2022vjx}, we consider the following potential depths: \( U_\Sigma = 30\,\rm MeV \) and \( U_\Xi = -4\,\rm MeV \). These potentials are in good agreement with those extracted from LQCD for hyperon interactions \cite{hyperon_lattice}. The hyperon masses are taken as \( m_\Lambda = 1116~{\rm MeV} \), \( m_\Sigma = 1193~{\rm MeV} \), \( m_\Xi = 1318~{\rm MeV} \) while \( M = 939~\text{MeV} \) and \( m_{\phi} = 1020~\text{MeV} \). The full Lagrangian for the baryon octet can be found in \cite{Biesdorf:2023icx}.

The hyperon-meson couplings according to SU(6) are:

\begin{equation}
    \dfrac{g_{\Lambda\omega}}{g_{N\omega}} = \dfrac{2}{3},\quad\quad  
    \dfrac{g_{\Sigma\omega}}{g_{N\omega}} = \dfrac{2}{3}, \quad{\rm and}\quad 
    \dfrac{g_{\Xi\omega}}{g_{N\omega}} = \dfrac{1}{3},
\end{equation}
for hyperon-\(\omega\) meson couplings.

\begin{equation}
    \dfrac{g_{\Lambda\phi}}{g_{N\omega}} =  \dfrac{-\sqrt{2}   }{3}, \quad
    \dfrac{g_{\Sigma\phi}}{g_{N\omega}} =  \dfrac{-\sqrt{2}}{3}, \quad{\rm and}\quad 
    \dfrac{g_{\Xi\phi}}{g_{N\omega}} =  \dfrac{-2\sqrt{2}}{3},
\end{equation}
for hyperon-\(\phi\) meson couplings.

\begin{equation}
    \dfrac{g_{\Lambda\rho}}{g_{N\rho}} = 0, \quad\quad 
    \dfrac{g_{\Sigma\rho}}{g_{N\rho}} = 2, \quad{\rm and}\quad 
    \dfrac{g_{\Xi\rho}}{g_{N\rho}} = 1.
\end{equation}
for hyperon-\(\rho\) meson couplings.

While those based on potential depths:

\begin{equation}
    \dfrac{g_{\Lambda\sigma}}{g_{N\sigma}} = 0.610, \quad\quad 
    \dfrac{g_{\Sigma\sigma}}{g_{N\sigma}} = 0.406, \quad{\rm and}\quad 
    \dfrac{g_{\Xi\sigma}}{g_{N\sigma}} = 0.269.
\end{equation}
for hyperon-\(\sigma\) meson couplings. One can note that \( g_{N\phi} = 0 \).

The particle fraction and speed of sound as a function of the baryon number density for the $EL3\omega\rho Y$ parametrization are given by

\begin{figure}[H]
  \centering
     \begin{subfigure}{0.5\textwidth}
         \centering
         \includegraphics[scale=0.6]{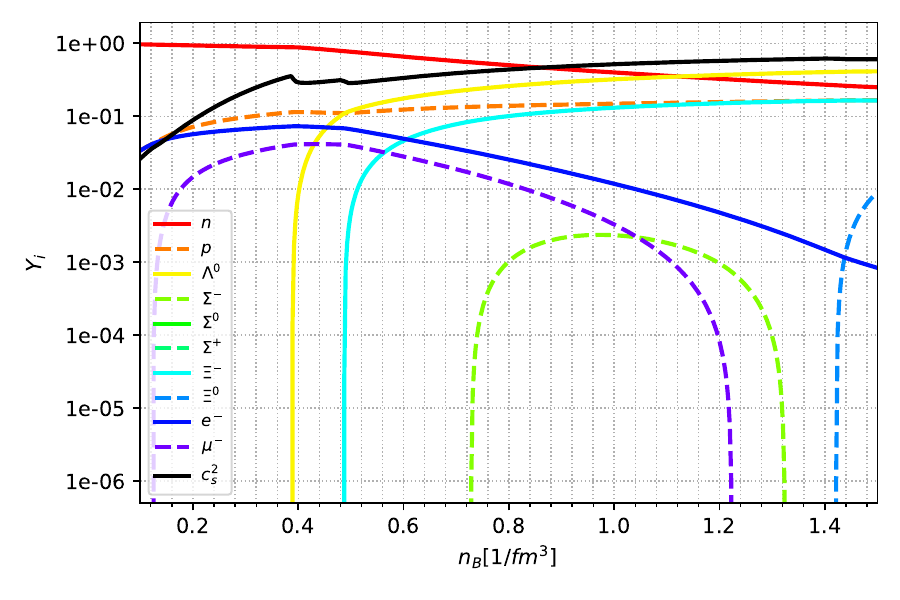}
     \end{subfigure}
        \caption{Particle fraction and speed of sound as a function of the baryon number density for the $EL3\omega\rho Y$ parametrization.}
         \label{fig_mb}         
\end{figure}
The appearance of $\Lambda^{0}$ and $\Xi^-$ significantly reduces the speed of sound ($c_{s}^2$), while the appearance of $\Xi^0$ at high densities leads to a slight reduction in the speed of sound.

\bibliographystyle{apsrev4-2}
\bibliography{ref}

@INPROCEEDINGS{2012IJMPS..18..105V,
       author = {{V{\'a}squez Flores}, C. and {Lenzi}, C.~H. and {Lugones}, G.},
        title = "{Radial Pulsations of Hybrid Neutron Stars}",
    booktitle = {International Journal of Modern Physics Conference Series},
         year = 2012,
       series = {International Journal of Modern Physics Conference Series},
       volume = {18},
        month = jan,
        pages = {105-108},
          doi = {10.1142/S201019451200829X},
       adsurl = {https://ui.adsabs.harvard.edu/abs/2012IJMPS..18..105V}
}

@article{Zhu:2018ona,
    author = "Zhu, Zhen-Yu and Zhou, En-Ping and Li, Ang",
    title = "{Neutron Star Equation of State from the Quark Level in Light of GW170817}",
    eprint = "1802.05510",
    archivePrefix = "arXiv",
    primaryClass = "nucl-th",
    doi = "10.3847/1538-4357/aacc28",
    journal = "Astrophys. J.",
    volume = "862",
    number = "2",
    pages = "98",
    year = "2018"
}

@article{Podder:2023dey,
    author = "Podder, Soumen and Pal, Suman and Sen, Debashree and Chaudhuri, Gargi",
    title = "{Constraints on density dependent MIT bag model parameters for quark and hybrid stars}",
    eprint = "2311.08962",
    archivePrefix = "arXiv",
    primaryClass = "nucl-th",
    doi = "10.1016/j.nuclphysa.2023.122796",
    journal = "Nucl. Phys. A",
    volume = "1042",
    pages = "122796",
    year = "2024"
}

@article{kumar,
  title = {Hybrid stars are compatible with recent astrophysical observations},
  author = {Kumar, Anil and Thapa, Vivek Baruah and Sinha, Monika},
  journal = {Phys. Rev. D},
  volume = {107},
  issue = {6},
  pages = {063024},
  numpages = {11},
  year = {2023},
  month = {Mar},
  publisher = {American Physical Society},
  doi = {10.1103/PhysRevD.107.063024},
  url = {https://link.aps.org/doi/10.1103/PhysRevD.107.063024}
}

@article{jia,
  title = {Relativistic hybrid stars with sequential first-order phase transitions and heavy-baryon envelopes},
  author = {Li, Jia Jie and Sedrakian, Armen and Alford, Mark},
  journal = {Phys. Rev. D},
  volume = {101},
  issue = {6},
  pages = {063022},
  numpages = {12},
  year = {2020},
  month = {Mar},
  publisher = {American Physical Society},
  doi = {10.1103/PhysRevD.101.063022},
  url = {https://link.aps.org/doi/10.1103/PhysRevD.101.063022}
}

@article{YAGI20171,
title = {Approximate universal relations for neutron stars and quark stars},
journal = {Physics Reports},
volume = {681},
pages = {1-72},
year = {2017},
note = {Approximate Universal Relations for Neutron Stars and Quark Stars},
issn = {0370-1573},
doi = {https://doi.org/10.1016/j.physrep.2017.03.002},
url = {https://www.sciencedirect.com/science/article/pii/S0370157317300492},
author = {Kent Yagi and Nicolás Yunes}
}

@article{Raithel_2018,
doi = {10.3847/2041-8213/aabcbf},
url = {https://doi.org/10.3847/2041-8213/aabcbf},
year = {2018},
month = {apr},
publisher = {The American Astronomical Society},
volume = {857},
number = {2},
pages = {L23},
author = {Raithel, Carolyn A. and Özel, Feryal and Psaltis, Dimitrios},
title = {Tidal Deformability from GW170817 as a Direct Probe of the Neutron Star Radius},
journal = {The Astrophysical Journal Letters}
}

@article{Kopp:2025hgp,
    author = {K{\"o}pp, F{\'a}bio and Horvath, Jorge Ernesto and Hadjimichef, Dimiter and Vasconcellos, C{\'e}sar A. Zen},
    title = "{Investigating Phase Transitions in Hybrid Stars Using the vMIT Bag Model and the QHD-II Model}",
    doi = "10.1002/asna.20250006",
    journal = "Astron. Nachr.",
    volume = "346",
    number = "3-4",
    pages = "e20250006",
    year = "2025"
}

@article{Heiselberg:2000dn,
    author = "Heiselberg, Henning and Pandharipande, Vijay",
    title = "{Recent progress in neutron star theory}",
    eprint = "astro-ph/0003276",
    archivePrefix = "arXiv",
    doi = "10.1146/annurev.nucl.50.1.481",
    journal = "Ann. Rev. Nucl. Part. Sci.",
    volume = "50",
    pages = "481--524",
    year = "2000"
}

@article{Boguta1977,
  author       = {J. Boguta and A. R. Bodmer},
  title        = {Relativistic calculation of nuclear matter and the nuclear surface},
  journal      = {Nuclear Physics A},
  volume       = {292},
  pages        = {413--428},
  year         = {1977},
  doi          = {10.1016/0375-9474(77)90626-1}
}

@article{Albino:2024ymc,
    author = "Albino, Milena and Malik, Tuhin and Ferreira, M\'arcio and Provid\^encia, Constan\c{c}a",
    title = "{Hybrid star properties with the NJL and mean field approximation of QCD models: A Bayesian approach}",
    eprint = "2406.15337",
    archivePrefix = "arXiv",
    primaryClass = "nucl-th",
    doi = "10.1103/PhysRevD.110.083037",
    journal = "Phys. Rev. D",
    volume = "110",
    number = "8",
    pages = "083037",
    year = "2024"
}

@article{Lopes:2022vjx,
    author = "Lopes, Luiz L. and Marquez, Kauan D. and Menezes, D\'ebora P.",
    title = "{Baryon coupling scheme in a unified SU(3) and SU(6) symmetry formalism}",
    eprint = "2211.17153",
    archivePrefix = "arXiv",
    primaryClass = "hep-ph",
    doi = "10.1103/PhysRevD.107.036011",
    journal = "Phys. Rev. D",
    volume = "107",
    number = "3",
    pages = "036011",
    year = "2023"
}

@article{hyperon_lattice,
  author  = {Inoue, T. and {HAL QCD Collaboration}},
  title   = {Hyperon Forces from QCD and Their Applications},
  journal = {JPS Conference Proceedings},
  volume  = {26},
  pages   = {023018},
  year    = {2019},
  doi     = {10.7566/JPSCP.26.023018},
  url     = {https://journals.jps.jp/doi/10.7566/JPSCP.26.023018}
}

@article{Lopes:2020rqn,
    author = "Lopes, Luiz L. and Menezes, D\'ebora P.",
    title = "{Broken SU(6) symmetry and massive hybrid stars}",
    eprint = "2004.07909",
    archivePrefix = "arXiv",
    primaryClass = "astro-ph.HE",
    doi = "10.1016/j.nuclphysa.2021.122171",
    journal = "Nucl. Phys. A",
    volume = "1009",
    pages = "122171",
    year = "2021"
}

@article{RevModPhys.38.215,
    author = "Pais, A.",
    title = "{Dynamical Symmetry in Particle Physics}",
    doi = "10.1103/revmodphys.38.215",
    journal = "Rev. Mod. Phys.",
    volume = "38",
    number = "2",
    pages = "215--255",
    year = "1966"
}

@article{matplotlib,
  author  = {Caswell, Thomas A. and Droettboom, Michael and Lee, Antony and et al.},
  title   = {Matplotlib: Visualization with Python},
  journal = {Journal of Open Source Software},
  year    = {2021},
  volume  = {6},
  number  = {60},
  pages   = {3027},
  doi     = {10.21105/joss.03027}
}

@article{Lopes:2021zfe,
    author = "Lopes, Luiz L.",
    title = "{Hyperonic neutron stars: reconciliation between nuclear properties and NICER and LIGO/VIRGO results}",
    eprint = "2107.02245",
    archivePrefix = "arXiv",
    primaryClass = "hep-ph",
    doi = "10.1088/1572-9494/ac2297",
    journal = "Commun. Theor. Phys.",
    volume = "74",
    number = "1",
    pages = "015302",
    year = "2022"
}

@article{Lopes:2021yga,
    author = "Lopes, Luiz L. and Menezes, Debora P.",
    title = "{On the Nature of the Mass-gap Object in the GW190814 Event}",
    eprint = "2111.02247",
    archivePrefix = "arXiv",
    primaryClass = "hep-ph",
    reportNumber = "APS/123-QED",
    doi = "10.3847/1538-4357/ac81c4",
    journal = "Astrophys. J.",
    volume = "936",
    number = "1",
    pages = "41",
    year = "2022"
}

@article{Mariani:2024gqi,
    author = "Mariani, Mauro and Ranea-Sandoval, Ignacio F. and Lugones, Germ\'an and Orsaria, Milva G.",
    title = "{Could a slow stable hybrid star explain the central compact object in HESS J1731-347?}",
    eprint = "2407.06347",
    archivePrefix = "arXiv",
    primaryClass = "astro-ph.HE",
    doi = "10.1103/PhysRevD.110.043026",
    journal = "Phys. Rev. D",
    volume = "110",
    number = "4",
    pages = "043026",
    year = "2024"
}

@article{Hebeler:2009iv,
    author = "Hebeler, K. and Schwenk, A.",
    title = "{Chiral three-nucleon forces and neutron matter}",
    eprint = "0911.0483",
    archivePrefix = "arXiv",
    primaryClass = "nucl-th",
    doi = "10.1103/PhysRevC.82.014314",
    journal = "Phys. Rev. C",
    volume = "82",
    pages = "014314",
    year = "2010"
}

@article{PhysRevD.81.105021,
  title = {Cold quark matter},
  author = {Kurkela, Aleksi and Romatschke, Paul and Vuorinen, Aleksi},
  journal = {Phys. Rev. D},
  volume = {81},
  issue = {10},
  pages = {105021},
  numpages = {32},
  year = {2010},
  month = {May},
  publisher = {American Physical Society},
  doi = {10.1103/PhysRevD.81.105021},
  url = {https://link.aps.org/doi/10.1103/PhysRevD.81.105021}
}

@ARTICLE{2018ApJ...860...12P,
       author = {{Pereira}, Jonas P. and {Flores}, C{\'e}sar V. and {Lugones}, Germ{\'a}n},
        title = "{Phase Transition Effects on the Dynamical Stability of Hybrid Neutron Stars}",
      journal = {\apj},
         year = 2018,
        month = jun,
       volume = {860},
       number = {1},
          eid = {12},
        pages = {12},
          doi = {10.3847/1538-4357/aabfbf},
archivePrefix = {arXiv},
       eprint = {1706.09371},
 primaryClass = {gr-qc},
       adsurl = {https://ui.adsabs.harvard.edu/abs/2018ApJ...860...12P}
}

@article{PhysRevD.109.083021,
  title = {Gravitational wave asteroseismology of dark matter hadronic stars},
  author = {Flores, C\'esar V. and Lenzi, C. H. and Dutra, M. and Louren\ifmmode \mbox{\c{c}}\else \c{c}\fi{}o, O. and Arba\~nil, Jos\'e D. V.},
  journal = {Phys. Rev. D},
  volume = {109},
  issue = {8},
  pages = {083021},
  numpages = {12},
  year = {2024},
  month = {Apr},
  publisher = {American Physical Society},
  doi = {10.1103/PhysRevD.109.083021},
  url = {https://link.aps.org/doi/10.1103/PhysRevD.109.083021}
}

@article{science.1236462,
author = {Kent Yagi  and Nicolás Yunes },
title = {I-Love-Q: Unexpected Universal Relations for Neutron Stars and Quark Stars},
journal = {Science},
volume = {341},
number = {6144},
pages = {365-368},
year = {2013},
doi = {10.1126/science.1236462},
URL = {https://www.science.org/doi/abs/10.1126/science.1236462},
eprint = {https://www.science.org/doi/pdf/10.1126/science.1236462}}

@article{Yagi_2016,
doi = {10.1088/0264-9381/33/13/13LT01},
url = {https://dx.doi.org/10.1088/0264-9381/33/13/13LT01},
year = {2016},
month = {jun},
publisher = {IOP Publishing},
volume = {33},
number = {13},
pages = {13LT01},
author = {Kent Yagi and Nicolás Yunes},
title = {Binary Love relations},
journal = {Classical and Quantum Gravity}
}

@article{Wiringa:1988tp,
  author = {Wiringa, R. B. and Fiks, V. and Fabrocini, A.},
  title = {Equation of state of dense nuclear matter},
  journal = {Phys. Rev. C},
  volume = {38},
  pages = {1010--1037},
  year = {1988}
}

@article{Mueller:1996pm,
  author = {Mueller, H. and Serot, B. D.},
  title = {Relativistic mean-field theory and the high-density nuclear equation of state},
  journal = {Nucl. Phys. A},
  volume = {606},
  pages = {508--537},
  year = {1996},
  eprint = {nucl-th/9603037},
  archivePrefix = {arXiv}
}

@article{LS,
  author = {Lattimer, J. M. and Swesty, F. Douglas},
  title = {A Generalized equation of state for hot, dense matter},
  journal = {Nucl. Phys. A},
  volume = {535},
  pages = {331--376},
  year = {1991}
}

@article{APR,
  author = {Akmal, A. and Pandharipande, V. and Ravenhall, D.},
  title = {Equation of state of nucleon matter and neutron star structure},
  journal = {Phys. Rev. C},
  volume = {58},
  pages = {1804--1828},
  year = {1998},
  eprint = {nucl-th/9804027},
  archivePrefix = {arXiv}
}

@article{PhysRevD.81.123016,
  title = {Tidal deformability of neutron stars with realistic equations of state and their gravitational wave signatures in binary inspiral},
  author = {Hinderer, Tanja and Lackey, Benjamin D. and Lang, Ryan N. and Read, Jocelyn S.},
  journal = {Phys. Rev. D},
  volume = {81},
  issue = {12},
  pages = {123016},
  numpages = {12},
  year = {2010},
  month = {Jun},
  publisher = {American Physical Society},
  doi = {10.1103/PhysRevD.81.123016},
  url = {https://link.aps.org/doi/10.1103/PhysRevD.81.123016}
}

@article{Annala:2019puf,
    author = {Annala, Eemeli and Gorda, Tyler and Kurkela, Aleksi and N\"attil\"a, Joonas and Vuorinen, Aleksi},
    title = "{Evidence for quark-matter cores in massive neutron stars}",
    eprint = "1903.09121",
    archivePrefix = "arXiv",
    primaryClass = "astro-ph.HE",
    reportNumber = "CERN-TH-2019-031, HIP-2019-7/TH",
    doi = "10.1038/s41567-020-0914-9",
    journal = "Nature Phys.",
    volume = "16",
    number = "9",
    pages = "907--910",
    year = "2020"
}

@article{PhysRevD.80.084035,
  title = {Relativistic tidal properties of neutron stars},
  author = {Damour, Thibault and Nagar, Alessandro},
  journal = {Phys. Rev. D},
  volume = {80},
  issue = {8},
  pages = {084035},
  numpages = {21},
  year = {2009},
  month = {Oct},
  publisher = {American Physical Society},
  doi = {10.1103/PhysRevD.80.084035},
  url = {https://link.aps.org/doi/10.1103/PhysRevD.80.084035}
}

@article{deMouraSoares:2023cqq,
    author = "de Moura Soares, Bruno A. and Lenzi, C{\'e}sar H. and Louren{\c{c}}o, Odilon and Dutra, Mariana",
    title = "{Bayesian analysis of a relativistic hadronic model constrained by recent astrophysical observations}",
    eprint = "2308.14417",
    archivePrefix = "arXiv",
    primaryClass = "nucl-th",
    doi = "10.1093/mnras/stad2558",
    journal = "Mon. Not. Roy. Astron. Soc.",
    volume = "525",
    number = "3",
    pages = "4347--4357",
    year = "2023"
}

@article{Saes:2021fzr,
    author = "Saes, Jayana A. and Mendes, Raissa F. P.",
    title = "{Equation-of-state-insensitive measure of neutron star stiffness}",
    eprint = "2109.11571",
    archivePrefix = "arXiv",
    primaryClass = "gr-qc",
    doi = "10.1103/PhysRevD.106.043027",
    journal = "Phys. Rev. D",
    volume = "106",
    number = "4",
    pages = "043027",
    year = "2022"
}

@article{Bodmer:1971we,
    author = "Bodmer, A. R.",
    title = "{Collapsed nuclei}",
    doi = "10.1103/PhysRevD.4.1601",
    journal = "Phys. Rev. D",
    volume = "4",
    pages = "1601--1606",
    year = "1971"
}

@article{Graeff_2019,
doi = {10.1088/1475-7516/2019/01/024},
url = {https://dx.doi.org/10.1088/1475-7516/2019/01/024},
year = {2019},
month = {jan},
publisher = {},
volume = {2019},
number = {01},
pages = {024},
author = {Clebson A. Graeff and Marcelo D. Alloy and Kauan D. Marquez and Constança Providência and Débora P. Menezes},
title = {Hadron-quark phase transition: the QCD phase diagram and stellar conversion},
journal = {Journal of Cosmology and Astroparticle Physics}
}

@article{Biesdorf:2023icx,
    author = "Biesdorf, Carline and Menezes, Debora P. and Lopes, Luiz L.",
    title = "{QCD Phase Diagrams via QHD and MIT-Based Models}",
    eprint = "2303.10734",
    archivePrefix = "arXiv",
    primaryClass = "hep-ph",
    doi = "10.1007/s13538-023-01348-z",
    journal = "Braz. J. Phys.",
    volume = "53",
    number = "5",
    pages = "137",
    year = "2023"
}

@article{kau17,
doi = {10.1088/1475-7516/2017/12/028},
url = {https://dx.doi.org/10.1088/1475-7516/2017/12/028},
year = {2017},
month = {dec},
publisher = {},
volume = {2017},
number = {12},
pages = {028},
author = {Marquez, Kauan D. and Menezes, Débora P.},
title = {Phase transition in compact stars: nucleation mechanism and γ-ray bursts revisited},
journal = {Journal of Cosmology and Astroparticle Physics}
}

@article{Terazawa:1989iw,
    author = "Terazawa, Hidezumi",
    title = "{Superhypernuclei in the Quark Shell Model}",
    reportNumber = "INS-732",
    doi = "10.1143/JPSJ.58.3555",
    journal = "J. Phys. Soc. Jap.",
    volume = "58",
    pages = "3555--3563",
    year = "1989"
}

@article{Witten:1984rs,
    author = "Witten, Edward",
    title = "{Cosmic Separation of Phases}",
    reportNumber = "PRINT-84-0400 (IAS,PRINCETON)",
    doi = "10.1103/PhysRevD.30.272",
    journal = "Phys. Rev. D",
    volume = "30",
    pages = "272--285",
    year = "1984"
}

@article{tov,
    author = "Oppenheimer, J. R. and Volkoff, G. M.",
    title = "{On massive neutron cores}",
    doi = "10.1103/PhysRev.55.374",
    journal = "Phys. Rev.",
    volume = "55",
    pages = "374--381",
    year = "1939"
}

@article{Huang:2023grj,
    author = "Huang, Chun and Raaijmakers, Geert and Watts, Anna L. and Tolos, Laura and Provid\^encia, Constan\c{c}a",
    title = "{Constraining a relativistic mean field model using neutron star mass\textendash{}radius measurements I: nucleonic models}",
    eprint = "2303.17518",
    archivePrefix = "arXiv",
    primaryClass = "astro-ph.HE",
    doi = "10.1093/mnras/stae844",
    journal = "Mon. Not. Roy. Astron. Soc.",
    volume = "529",
    number = "4",
    pages = "4650--4665",
    year = "2024"
}

@article{Malik:2022zol,
    author = "Malik, Tuhin and Ferreira, M\'arcio and Agrawal, B. K. and Provid\^encia, Constan\c{c}a",
    title = "{Relativistic Description of Dense Matter Equation of State and Compatibility with Neutron Star Observables: A Bayesian Approach}",
    eprint = "2201.12552",
    archivePrefix = "arXiv",
    primaryClass = "nucl-th",
    doi = "10.3847/1538-4357/ac5d3c",
    journal = "Astrophys. J.",
    volume = "930",
    number = "1",
    pages = "17",
    year = "2022"
}

@article{Wesolowski:2015fqa,
    author = "Wesolowski, S. and Klco, N. and Furnstahl, R. J. and Phillips, D. R. and Thapaliya, A.",
    title = "{Bayesian parameter estimation for effective field theories}",
    eprint = "1511.03618",
    archivePrefix = "arXiv",
    primaryClass = "nucl-th",
    doi = "10.1088/0954-3899/43/7/074001",
    journal = "J. Phys. G",
    volume = "43",
    number = "7",
    pages = "074001",
    year = "2016"
}

@article{Imam:2024gfh,
    author = "Imam, Sk Md Adil and Malik, Tuhin and Provid\^encia, Constan\c{c}a and Agrawal, B. K.",
    title = "{Implications of comprehensive nuclear and astrophysics data on the equations of state of neutron star matter}",
    eprint = "2401.06018",
    archivePrefix = "arXiv",
    primaryClass = "nucl-th",
    doi = "10.1103/PhysRevD.109.103025",
    journal = "Phys. Rev. D",
    volume = "109",
    number = "10",
    pages = "103025",
    year = "2024"
}

@article{Char:2023fue,
    author = "Char, Prasanta and Mondal, Chiranjib and Gulminelli, Francesca and Oertel, Micaela",
    title = "{Generalized description of neutron star matter with a nucleonic relativistic density functional}",
    eprint = "2307.12364",
    archivePrefix = "arXiv",
    primaryClass = "nucl-th",
    doi = "10.1103/PhysRevD.108.103045",
    journal = "Phys. Rev. D",
    volume = "108",
    number = "10",
    pages = "103045",
    year = "2023"
}

@article{Ashton:2018jfp,
    author = "Ashton, Gregory and others",
    title = "{BILBY: A user-friendly Bayesian inference library for gravitational-wave astronomy}",
    eprint = "1811.02042",
    archivePrefix = "arXiv",
    primaryClass = "astro-ph.IM",
    doi = "10.3847/1538-4365/ab06fc",
    journal = "Astrophys. J. Suppl.",
    volume = "241",
    number = "2",
    pages = "27",
    year = "2019"
}

@article{Furnstahl:2015rha,
    author = "Furnstahl, R. J. and Klco, N. and Phillips, D. R. and Wesolowski, S.",
    title = "{Quantifying truncation errors in effective field theory}",
    eprint = "1506.01343",
    archivePrefix = "arXiv",
    primaryClass = "nucl-th",
    doi = "10.1103/PhysRevC.92.024005",
    journal = "Phys. Rev. C",
    volume = "92",
    number = "2",
    pages = "024005",
    year = "2015"
}

@article{bps,
    author = {Baym, Gordon and Pethick, C. and Sutherland, P.},
    title = {The Ground State of Matter at High Densities: Equation of State and Stellar Models},
    journal = {The Astrophysical Journal},
    year = {1971},
    volume = {170},
    pages = {299-317},
    doi = {10.1086/151216}
}

@article{Bayes1763,
  author = {Bayes, Thomas},
  title = {An Essay towards Solving a Problem in the Doctrine of Chances},
  journal = {Philosophical Transactions of the Royal Society of London},
  volume = {53},
  pages = {370--418},
  year = {1763},
  doi = {10.1098/rstl.1763.0053},
  url = {https://doi.org/10.1098/rstl.1763.0053}
}

@article{Chodos1974,
  author = {Chodos, A. and Jaffe, R. L. and Johnson, K. and Thorn, C. B. and Weisskopf, V. F.},
  title = {A New Extended Model of Hadrons},
  journal = {Physical Review D},
  volume = {9},
  number = {12},
  pages = {3471-3495},
  year = {1974},
  doi = {10.1103/PhysRevD.9.3471},
  url = {https://doi.org/10.1103/PhysRevD.9.3471}
}

@article{Fukushima_2011,
doi = {10.1088/0034-4885/74/1/014001},
url = {https://dx.doi.org/10.1088/0034-4885/74/1/014001},
year = {2010},
month = {dec},
publisher = {},
volume = {74},
number = {1},
pages = {014001},
author = {Kenji Fukushima and Tetsuo Hatsuda},
title = {The phase diagram of dense QCD},
journal = {Reports on Progress in Physics}
}

@article{Voskresensky:2002hu,
    author = "Voskresensky, D. N. and Yasuhira, M. and Tatsumi, T.",
    title = "{Charge screening at first order phase transitions and hadron quark mixed phase}",
    eprint = "nucl-th/0208067",
    archivePrefix = "arXiv",
    doi = "10.1016/S0375-9474(03)01313-7",
    journal = "Nucl. Phys. A",
    volume = "723",
    pages = "291--339",
    year = "2003"
}

@article{Lugones:2016ytl,
    author = "Lugones, G. and Grunfeld, A. G.",
    title = "{Surface tension of highly magnetized degenerate quark matter}",
    eprint = "1610.05875",
    archivePrefix = "arXiv",
    primaryClass = "nucl-th",
    doi = "10.1103/PhysRevC.95.015804",
    journal = "Phys. Rev. C",
    volume = "95",
    number = "1",
    pages = "015804",
    year = "2017"
}

@article{Lugones:2013ema,
    author = "Lugones, G. and Grunfeld, A. G. and Al Ajmi, M.",
    title = "{Surface tension and curvature energy of quark matter in the Nambu-Jona-Lasinio model}",
    eprint = "1308.1452",
    archivePrefix = "arXiv",
    primaryClass = "hep-ph",
    doi = "10.1103/PhysRevC.88.045803",
    journal = "Phys. Rev. C",
    volume = "88",
    number = "4",
    pages = "045803",
    year = "2013"
}

@article{Pinto:2012aq,
    author = "Pinto, Marcus B. and Koch, Volker and Randrup, Jorgen",
    title = "{The Surface Tension of Quark Matter in a Geometrical Approach}",
    eprint = "1207.5186",
    archivePrefix = "arXiv",
    primaryClass = "hep-ph",
    doi = "10.1103/PhysRevC.86.025203",
    journal = "Phys. Rev. C",
    volume = "86",
    pages = "025203",
    year = "2012"
}

@article{Lugones:2018qgu,
    author = "Lugones, G. and Grunfeld, A. G.",
    title = "{Surface tension of hot and dense quark matter under strong magnetic fields}",
    eprint = "1811.09954",
    archivePrefix = "arXiv",
    primaryClass = "astro-ph.HE",
    doi = "10.1103/PhysRevC.99.035804",
    journal = "Phys. Rev. C",
    volume = "99",
    number = "3",
    pages = "035804",
    year = "2019"
}

@article{Maruyama:2007ey,
    author = "Maruyama, Toshiki and Chiba, Satoshi and Schulze, Hans-Josef and Tatsumi, Toshitaka",
    title = "{Hadron-quark mixed phase in hyperon stars}",
    eprint = "0708.3277",
    archivePrefix = "arXiv",
    primaryClass = "nucl-th",
    doi = "10.1103/PhysRevD.76.123015",
    journal = "Phys. Rev. D",
    volume = "76",
    pages = "123015",
    year = "2007"
}

@article{Laskos-Patkos:2023tlr,
    author = "Laskos-Patkos, P. and Koliogiannis, P. S. and Moustakidis, Ch. C.",
    title = "{Hybrid stars in light of the HESS J1731-347 remnant and the PREX-II experiment}",
    eprint = "2312.07113",
    archivePrefix = "arXiv",
    primaryClass = "astro-ph.HE",
    doi = "10.1103/PhysRevD.109.063017",
    journal = "Phys. Rev. D",
    volume = "109",
    number = "6",
    pages = "063017",
    year = "2024"
}

@article{Lopes:2020btp,
    author = "Lopes, Luiz L. and Biesdorf, Carline and Menezes, D. \'ebora P.",
    title = "{Modified MIT bag Models\textemdash{}part I: Thermodynamic consistency, stability windows and symmetry group}",
    eprint = "2005.13136",
    archivePrefix = "arXiv",
    primaryClass = "hep-ph",
    doi = "10.1088/1402-4896/abef34",
    journal = "Phys. Scripta",
    volume = "96",
    number = "6",
    pages = "065303",
    year = "2021"
}

@article{ForemanMackey2016,
  author = {Foreman-Mackey, Daniel},
  title = {corner.py: Scatterplot matrices in Python},
  journal = {The Journal of Open Source Software},
  volume = {1},
  number = {2},
  pages = {24},
  year = {2016},
  doi = {10.21105/joss.00024},
  url = {https://doi.org/10.21105/joss.00024}
}

@article{sophia_quark_c,
doi = {10.3847/1538-4357/abf355},
url = {https://dx.doi.org/10.3847/1538-4357/abf355},
year = {2021},
month = {may},
publisher = {The American Astronomical Society},
volume = {913},
number = {1},
pages = {27},
author = {Ang Li and Zhiqiang Miao and Sophia Han and Bing Zhang},
title = {Constraints on the Maximum Mass of Neutron Stars with a Quark Core from GW170817 and NICER PSR J0030+0451 Data},
journal = {The Astrophysical Journal}
}

@article{PhysRevLett.120.172703,
  title = {Gravitational-Wave Constraints on the Neutron-Star-Matter Equation of State},
  author = {Annala, Eemeli and Gorda, Tyler and Kurkela, Aleksi and Vuorinen, Aleksi},
  journal = {Phys. Rev. Lett.},
  volume = {120},
  issue = {17},
  pages = {172703},
  numpages = {5},
  year = {2018},
  month = {Apr},
  publisher = {American Physical Society},
  doi = {10.1103/PhysRevLett.120.172703},
  url = {https://link.aps.org/doi/10.1103/PhysRevLett.120.172703}
}

@article{Lopes2021c,
  author    = {Luiz L. Lopes and Carline Biesdorf and Débora P. Menezes},
  title     = {Hypermassive quark cores},
  journal   = {Monthly Notices of the Royal Astronomical Society},
  volume    = {512},
  number    = {4},
  pages     = {5110--5121},
  year      = {2022},
  doi       = {10.1093/mnras/stac793}
}

@article{dutra2014,
  author    = {Mariana Dutra and O. Lourenço and S. S. Avancini and B. V. Carlson and A. Delfino and D. P. Menezes and C. Providência and S. Typel and J. R. Stone},
  title     = {Relativistic mean-field hadronic models under nuclear matter constraints},
  journal   = {Physical Review C},
  volume    = {90},
  number    = {5},
  pages     = {055203},
  year      = {2014},
  doi       = {10.1103/PhysRevC.90.055203}
}

@article{Walecka,
  author    = {J. D. Walecka},
  title     = {A Theory of Highly Condensed Matter},
  journal   = {Annals of Physics},
  volume    = {83},
  number    = {2},
  pages     = {491--529},
  year      = {1974},
  doi       = {10.1016/0003-4916(74)90208-5}
}

@article{Nagata:2021ugx,
    author = "Nagata, Keitaro",
    title = "{Finite-density lattice QCD and sign problem: Current status and open problems}",
    eprint = "2108.12423",
    archivePrefix = "arXiv",
    primaryClass = "hep-lat",
    doi = "10.1016/j.ppnp.2022.103991",
    journal = "Prog. Part. Nucl. Phys.",
    volume = "127",
    pages = "103991",
    year = "2022"
}

@article{aoki2006order,
  title={The order of the quantum chromodynamics transition predicted by the standard model of particle physics},
  author={Aoki, Yasumichi and Endr{\H{o}}di, G and Fodor, Zolt{\'a}n and Katz, S{\'a}ndor D and Szab{\'o}, K{\'a}lm{\'a}n K},
  journal={Nature},
  volume={443},
  number={7112},
  pages={675--678},
  year={2006},
  publisher={Nature Publishing Group}
}

@article{bellwied2015qcd,
  title={The QCD phase diagram from analytic continuation},
  author={Bellwied, Rene and Bors{\'a}nyi, Szabolcs and Fodor, Zolt{\'a}n and G{\"u}nther, J and Katz, SD and Ratti, C and Szabo, KK},
  journal={Physics Letters B},
  volume={751},
  pages={559--564},
  year={2015},
  publisher={Elsevier}
}

@article{luostudy,
  title={{A Study of the Properties of the QCD Phase Diagram in High-Energy Nuclear Collisions}},
  author={Luo, Xiaofeng and Shi, Shusu and Xu, Nu and Zhang, Yifei},
  journal={Particles},
  volume={3},
  number={2},
  pages={278--307},
  year={2020},
  publisher={MDPI},
  doi={10.3390/particles3020022},
  url={https://www.mdpi.com/2571-712X/3/2/22}
}

@ARTICLE{1992PhRvD,
       author = {{Glendenning}, Norman K.},
        title = "{First-order phase transitions with more than one conserved charge: Consequences for neutron stars}",
      journal = {\prd},
         year = 1992,
        month = aug,
       volume = {46},
       number = {4},
        pages = {1274-1287},
          doi = {10.1103/PhysRevD.46.1274},
       adsurl = {https://ui.adsabs.harvard.edu/abs/1992PhRvD..46.1274G}
}

@article{bazavov2017skewness,
  title={Skewness and kurtosis of net baryon-number distributions at small values of the baryon chemical potential},
  author={Bazavov, A and Ding, H-T and Hegde, P and Kaczmarek, Olaf and Karsch, Frithjof and Laermann, Edwin and Mukherjee, Swagato and Ohno, Hiroshi and Petreczky, P and Rinaldi, E and others},
  journal={Physical Review D},
  volume={96},
  number={7},
  pages={074510},
  year={2017},
  publisher={APS}
}

@article{bazavov2017qcd,
  title={QCD equation of state to O ($\mu$ B 6) from lattice QCD},
  author={Bazavov, A and Ding, H-T and Hegde, P and Kaczmarek, Olaf and Karsch, Frithjof and Laermann, Edwin and Maezawa, Y and Mukherjee, Swagato and Ohno, H and Petreczky, P and others},
  journal={Physical Review D},
  volume={95},
  number={5},
  pages={054504},
  year={2017},
  publisher={APS}
}

@article{riley2021nicer,
  title = {{A NICER view of the massive pulsar PSR J0740+6620 informed by radio timing and XMM-Newton spectroscopy}},
  author = {Riley, T. E. and Watts, A. L. and Ray, P. S. and Bogdanov, S. and Guillot, S. and Morsink, S. M. and Bilous, A. V. and Arzoumanian, Z. and Choudhury, D. and Deneva, J. S. and others},
  journal = {The Astrophysical Journal Letters},
  volume = {918},
  number = {2},
  pages = {L27},
  year = {2021}
}

@article{LIGOScientific:2017vwq,
    author = "Abbott, B. P. and others",
    collaboration = "LIGO Scientific, Virgo",
    title = "{GW170817: Observation of Gravitational Waves from a Binary Neutron Star Inspiral}",
    eprint = "1710.05832",
    archivePrefix = "arXiv",
    primaryClass = "gr-qc",
    reportNumber = "LIGO-P170817",
    doi = "10.1103/PhysRevLett.119.161101",
    journal = "Phys. Rev. Lett.",
    volume = "119",
    number = "16",
    pages = "161101",
    year = "2017"
}

@article{Miller2019,
    author = "Miller, M. C. and others",
    title = "{PSR J0030+0451 Mass and Radius from $NICER$ Data and Implications for the Properties of Neutron Star Matter}",
    eprint = "1912.05705",
    archivePrefix = "arXiv",
    primaryClass = "astro-ph.HE",
    doi = "10.3847/2041-8213/ab50c5",
    journal = "Astrophys. J. Lett.",
    volume = "887",
    number = "1",
    pages = "L24",
    year = "2019"
}

@article{Miller2021,
    author = "Miller, M. C. and others",
    title = "{The Radius of PSR J0740+6620 from NICER and XMM-Newton Data}",
    eprint = "2105.06979",
    archivePrefix = "arXiv",
    primaryClass = "astro-ph.HE",
    doi = "10.3847/2041-8213/ac089b",
    journal = "Astrophys. J. Lett.",
    volume = "918",
    number = "2",
    pages = "L28",
    year = "2021"
}

@article{Imam:2021dbe,
    author = "Imam, Sk Md Adil and Patra, N. K. and Mondal, C. and Malik, Tuhin and Agrawal, B. K.",
    title = "{Bayesian reconstruction of nuclear matter parameters from the equation~of state of neutron star matter}",
    eprint = "2110.15776",
    archivePrefix = "arXiv",
    primaryClass = "nucl-th",
    doi = "10.1103/PhysRevC.105.015806",
    journal = "Phys. Rev. C",
    volume = "105",
    number = "1",
    pages = "015806",
    year = "2022"
}

@article{Malik:2022ilb,
    author = "Malik, Tuhin and Agrawal, B. K. and Provid\^encia, Constan\c{c}a",
    title = "{Inferring the nuclear symmetry energy at suprasaturation density from neutrino cooling}",
    eprint = "2206.15404",
    archivePrefix = "arXiv",
    primaryClass = "nucl-th",
    doi = "10.1103/PhysRevC.106.L042801",
    journal = "Phys. Rev. C",
    volume = "106",
    number = "4",
    pages = "L042801",
    year = "2022"
}

@article{Coughlin:2019kqf,
    author = "Coughlin, Michael W. and Dietrich, Tim",
    title = "{Can a black hole\textendash{}neutron star merger explain GW170817, AT2017gfo, and GRB170817A?}",
    eprint = "1901.06052",
    archivePrefix = "arXiv",
    primaryClass = "astro-ph.HE",
    doi = "10.1103/PhysRevD.100.043011",
    journal = "Phys. Rev. D",
    volume = "100",
    number = "4",
    pages = "043011",
    year = "2019"
}

@article{Abbott2018_GW170817_Radii,
  author        = {Abbott, B. P. and others},
  collaboration = {LIGO Scientific Collaboration and Virgo Collaboration},
  title         = {GW170817: Measurements of Neutron Star Radii and Equation of State},
  journal       = {Physical Review Letters},
  volume        = {121},
  number        = {16},
  pages         = {161101},
  year          = {2018},
  doi           = {10.1103/PhysRevLett.121.161101},
  eprint        = {1805.11581},
  archivePrefix = {arXiv},
  primaryClass  = {gr-qc}
}

@article{Trotta01032008,
author = {Roberto Trotta},
title = {Bayes in the sky: Bayesian inference and model selection in cosmology},
journal = {Contemporary Physics},
volume = {49},
number = {2},
pages = {71--104},
year = {2008},
publisher = {Taylor \& Francis},
doi = {10.1080/00107510802066753},


URL = { 
    
        https://doi.org/10.1080/00107510802066753
    
    

},
eprint = { 
    
        https://doi.org/10.1080/00107510802066753
    
    

}

}

@article{Albino2025puc,
  title = {Bayesian inference of hybrid stars with large quark cores},
  author = {Albino, Milena and Malik, Tuhin and Ferreira, M\'arcio and Provid\^encia, Constan\ifmmode \mbox{\c{c}}\else \c{c}\fi{}a},
  journal = {Phys. Rev. D},
  volume = {113},
  issue = {8},
  pages = {083019},
  numpages = {23},
  year = {2026},
  month = {Apr},
  publisher = {American Physical Society},
  doi = {10.1103/jrz4-zjq1},
  url = {https://link.aps.org/doi/10.1103/jrz4-zjq1}
}

@article{dynesty,
    author = {Speagle, Joshua S},
    title = {dynesty: a dynamic nested sampling package for estimating Bayesian posteriors and evidences},
    journal = {Monthly Notices of the Royal Astronomical Society},
    volume = {493},
    number = {3},
    pages = {3132-3158},
    year = {2020},
    month = {02},
    issn = {0035-8711},
    doi = {10.1093/mnras/staa278},
    url = {https://doi.org/10.1093/mnras/staa278},
    eprint = {https://academic.oup.com/mnras/article-pdf/493/3/3132/32890730/staa278.pdf},
}

@article{Vinciguerra:2023qxq,
    author = "Vinciguerra, Serena and others",
    title = "{An Updated Mass{\textendash}Radius Analysis of the 2017{\textendash}2018 NICER Data Set of PSR J0030+0451}",
    eprint = "2308.09469",
    archivePrefix = "arXiv",
    primaryClass = "astro-ph.HE",
    doi = "10.3847/1538-4357/acfb83",
    journal = "Astrophys. J.",
    volume = "961",
    number = "1",
    pages = "62",
    year = "2024"
}

@article{Salmi:2024aum,
    author = "Salmi, Tuomo and others",
    title = "{The Radius of the High-mass Pulsar PSR J0740+6620 with 3.6 yr of NICER Data}",
    eprint = "2406.14466",
    archivePrefix = "arXiv",
    primaryClass = "astro-ph.HE",
    doi = "10.3847/1538-4357/ad5f1f",
    journal = "Astrophys. J.",
    volume = "974",
    number = "2",
    pages = "294",
    year = "2024"
}

\end{document}